\documentclass[showpacs,showkeys,amsmath,amssymb,superscriptaddress,reprint, preprintnumbers,nofootinbib,notitlepage, groupdaddress,showabstract,aps,prl,twocolumn]{revtex4-1}

\pdfoutput=1 % if your are submitting a pdflatex (i.e. if you have
\usepackage[font=small]{caption}
\usepackage[utf8]{inputenc}
\usepackage[english]{babel}
\usepackage{bm}
\usepackage{graphicx}
\usepackage{dcolumn}
\usepackage{pbox}
\usepackage{epsfig}
\usepackage[dvipsnames]{xcolor}
\usepackage{slashed}
\usepackage{amssymb}
\usepackage{ mathrsfs }
\usepackage{color}
\usepackage[font=small]{caption}
\usepackage[font=small]{subcaption}
\usepackage{url}
\definecolor{DarkBlue}{rgb}{0.7, 0.4, 1} 
\definecolor{Blue}{rgb}{0, 0.8, 0} 
\definecolor{MyLightBlue}{rgb}{0.5,0.7,1.9}
\definecolor{MyGreen}{rgb}{0.0,0.2, 0.0}
\definecolor{MyBrickRed}{rgb}{0, 0.5, 0.2}
\RequirePackage{hyperref}
\hypersetup{colorlinks, citecolor=Blue,linkcolor=DarkBlue, urlcolor=Green}
\usepackage[normalem]{ulem}

\newcommand{\bea}{\begin{eqnarray}}
\newcommand{\eea}{\end{eqnarray}}

\makeatletter

\renewcommand\@makecaption[2]{%
  \par
  \vskip\abovecaptionskip
  \begingroup
  
   \small\rmfamily
    \begingroup
     \samepage
     \flushing
     \let\footnote\@footnotemark@gobble
     \@make@capt@title{#1}{#2}\par
    \endgroup
  \endgroup
  \vskip\belowcaptionskip
}
\makeatother

\DeclareUnicodeCharacter{2212}{-}
\newcommand{\Mzp}{M_{Z^\prime}}

\newcommand{\gs}{g_\star}
\newcommand{\gss}{g_{\star s}}
\newcommand{\tr}{\text{Tr}}

\begin{document}
%%%%%%%%%%%%%%%%%%%%%%%%%%%%%%
\title{Hunting for heavy $Z^\prime$ with IceCube neutrinos and gravitational waves} 
%%%%%%%%%%%%%%%%%%%%%
\author{Basabendu Barman}
\email{basabendu.b@srmap.edu.in}
\affiliation{\,\,Department of Physics, School of Engineering and Sciences, SRM University-AP, Amaravati 522240, India}
\author{Arindam Das}
\email{arindamdas@oia.hokudai.ac.jp}
\affiliation{\,\,Institute for the Advancement of Higher Education, Hokkaido University, Sapporo 060-0817, Japan}
\affiliation{Department of Physics, Hokkaido University, Sapporo 060-0810, Japan}
\author{Suruj Jyoti Das}
\email{surujjd@gmail.com}
\affiliation{\,\,Particle Theory  and Cosmology Group, Center for Theoretical Physics of the Universe, Institute for Basic Science (IBS),
 Daejeon, 34126, Korea}
\author{Marco Merchand}
\email{marcomm@kth.se}
\affiliation{\,\,KTH Royal Institute of Technology, Department of Physics, SE-10691 Stockholm, Sweden}
\affiliation{\,\,The Oskar Klein Centre for Cosmoparticle Physics, AlbaNova University Centre, SE-10691 Stockholm, Sweden}
\preprint{CTPU-PTC-25-06}
%%%%%%%%%%%%%%%%%%%%%%%%%%%
\begin{abstract}   
In the minimal gauged  B$-$L extension of the Standard Model, we demonstrate that PeV-scale dark matter (DM) and the baryon asymmetry of the Universe (BAU) can be simultaneously explained through the three right-handed neutrinos (RHNs) present in the theory. The DM candidate undergoes rare decay into light neutrinos, providing an explanation for the observed IceCube events, while the other two RHNs generate the BAU via leptogenesis. The breaking of gauge symmetry gives rise to detectable gravitational waves (GWs) from decaying cosmic strings (CS), making this framework testable at several future GW detectors—despite being beyond the reach of conventional collider experiments due to the extremely weak gauge coupling. The symmetry-breaking scale establishes a connection between particle masses, couplings, and the GW spectrum, offering a unified and predictive scenario. 
\end{abstract}
%%%%%%%%%%%%%%%%%%%%%%%%%%%
\maketitle
%%%%%%%%%%%%%%%%%%%%%%%%%%%%%%%%%%%%%%%%%%%%%%%%%%
\noindent

%%%%%%%%main text%%%%%%%%%%%%%%%%%%%%%%%%%%%%%%%%%
{\textbf{Introduction}.--}
Neutrino oscillation experiments have provided evidence for tiny neutrino masses and flavor mixing over time \cite{ParticleDataGroup:2024cfk}. The PLANCK \cite{Planck:2018vyg} observations estimate the baryon asymmetry of the universe (BAU), defined as the ratio of the baryon-antibaryon number density difference $(n_B - n_{\bar{B}})$ to the entropy density $(s)$, to be $Y_B^o \simeq 8.75 \times 10^{-11}$~\cite{Planck:2018vyg}. Furthermore, astrophysical and cosmological observations strongly support the existence of dark matter (DM), with its relic abundance measured as $\Omega h^2 \simeq 0.12$~\cite{Bertone:2016nfn,deSwart:2017heh,Planck:2018vyg}. These phenomena remain unexplained within the Standard Model (SM), necessitating exploration of beyond Standard Model (BSM) frameworks to uncover their fundamental origins.

In the recent past, the IceCube neutrino observatory reported the detection of three PeV neutrinos, roughly $3\sigma$ excess above the expected background rates~\cite{IceCube:2013cdw, IceCube:2013low, IceCube:2014stg,IceCube:2015gsk,IceCube:2015qii,Li:2025tqf}. These highest-energy events correspond to deposited energies of 1.04 PeV, 1.14 PeV and 2.0 PeV, respectively. Although the origin of these very high energy events is still unclear, it has been shown that such events could be originated from decays of superheavy DM ~\cite{Bai:2013nga,Higaki:2014dwa,Esmaili:2014rma,Murase:2015gea,Dudas:2018npp,Cohen:2016uyg,Rott:2014kfa}. The neutrino energy spectrum presents a high-energy cut-off at half of the DM mass~\cite{Higaki:2014dwa,Esmaili:2014rma} if two body decays including one neutrino are present. Moreover, the IceCube spectrum sets a lower bound on the DM lifetime $\tau_{\rm DM}\simeq \mathcal{O}(10^{28})$~s~\cite{Esmaili:2014rma,Arguelles:2022nbl}, which is largely model-independent and significantly exceeds the age of the Universe. 

On the other hand, current observations of Gravitational Waves (GWs)~\cite{LIGOScientific:2021nrg,Caldwell:2022qsj,NANOGrav:2023gor,NANOGrav:2023hvm,Xu:2023wog,EPTA:2023fyk,LISACosmologyWorkingGroup:2022jok}, have opened a complementary avenue to test BSM physics. For example, if symmetries are broken spontaneously at very high temperature then topological defects, such as cosmic strings and domain walls, may appear in the early stages of the universe~\cite{Nielsen:1973cs,Kibble:1976sj} and the system of these defects can be considered as a prominent source of a GW background, while the scale of symmetry breaking can be associated with the scale of new physics.

To explore these aspects within a simple yet elegant framework, we consider an extension of the SM featuring an anomaly-free  $U(1)_{\rm{B-L}}$ gauge group~\cite{Davidson:1978pm,Marshak:1979fm}, incorporating three generations of SM-singlet right-handed neutrinos (RHNs) and an SM-singlet scalar which acquires a non-zero vacuum expectation value (VEV) resulting in the breaking of the $U(1)_{\rm{B-L}}$ symmetry. As a result, the Majorana masses for the RHNs are generated, which in turn  induce tiny masses and flavor mixing for the observed left-handed neutrinos via the seesaw mechanism~\cite{Minkowski:1977sc,Yanagida:1979as,Gell-Mann:1979vob,Mohapatra:1979ia,Schechter:1980gr}. This setup naturally explains the BAU via vanilla leptogenesis if at least two RHN generations contribute to the asymmetry generation, while the third, lightest RHN can serve as a PeV-scale decaying DM candidate, addressing IceCube high-energy neutrino events through the freeze-in mechanism. Additionally, the $U(1)_{\rm{B-L}}$  breaking gives rise to one-dimensional topological defects—cosmic strings (CS)—characterized by a string tension $G\mu\sim BG v^2_\Phi$, where $v_\Phi$ is the VEV of the $U(1)_{\rm{B-L}}$ symmetry breaking singlet scalar and $B\sim 0.1$~\cite{Babul:1987me,Vilenkin:2000jqa,Dror:2019syi}. This minimal scenario, therefore, offers a unified explanation for (i) PeV-scale decaying DM linked to IceCube events, (ii) baryogenesis via leptogenesis and (iii) GW from CS which could be tested at future GW detectors. We thus constrain heavy neutral gauge boson mass beyond 1 TeV, with tiny gauge coupling $(\lesssim\mathcal{O}(10^{-5}))$, which lies beyond the reach of high energy collider experiments, that typically provide constraints on gauge couplings around $\mathcal{O}(10^{-2})$ for the B$-$L scenario at the LHC \cite{Das:2021esm}. At this stage it is worth pointing out that the IceCube events may also be accounted for by the high-energy neutrino flux originating from evaporating primordial black holes (PBHs)~\cite{Lunardini:2019zob,Bernal:2022swt,Wu:2024uxa,Chianese:2024rsn,Zantedeschi:2024ram,Baker:2025cff}. The central focus of the present work, however, is to establish a connection with the mechanism responsible for generating light neutrino masses—an aspect that unequivocally requires physics beyond the Standard Model (BSM). Rather than direct neutrino emission from PBHs, one could also consider scenarios where PBHs emit heavy BSM particles (such as heavy leptons), which subsequently decay into neutrinos. This indirect production channel has been widely explored in the literature and necessitates the inclusion of BSM fields such as right handed neutrinos (see, for example, Refs.~\cite{Fujita:2014hha,Datta:2020bht,Barman:2021ost,Bernal:2022pue}). In contrast, our approach does not rely on PBHs as the source, but considers the thermal SM radiation bath to be responsible for producing heavy neutral leptons (RHNs in present case), which ultimately lead to the observed neutrino signatures.
\\

%%%%%%%%%%%%%%%%%%%
\noindent
{\textbf{The framework}--}
%%%%%%%%%%%%%%%%%%%%%%%%%%%%%%%%%%%%%%%%%%%%%%%%
Under the $\text{SM}\otimes U(1)_{\rm B-L}$ gauge symmetry, the SM quark fields transform as $q_L^i=\{3,2,\frac{1}{6}, \frac{1}{3}\}$, $u_R^i=\{3,1,\frac{2}{3}, \frac{1}{3}\}$, $d_R^i=\{3,1,-\frac{1}{3}, \frac{1}{3}\}$, respectively.
The SM lepton fields transform as $\ell_L^i=\{1,2,-\frac{1}{2}, -1\}$, $e_R^i=\{1,1,-1, -1\}$, respectively, while the SM Higgs field $H=\{1,2,\frac{1}{2}, 0\}$. 
We introduce three SM-singlet RHNs to cancel gauge and mixed gauge-gravity anomalies which transform as $N_R^{i}=\{1,1,0,-1\}$ with $i=1, 2, 3$ and one SM-singlet $U(1)_{\rm{B-L}}$ scalar which transforms as $\Phi=\{1,1,0, 2\}$. The relevant Yukawa interactions read,
\bea
{\cal L} &\supset& - Y_{\nu_{\alpha \beta}} \overline{\ell_L^\alpha} \tilde{H}\,N_R^\beta- \frac{1}{2}Y_{N_\alpha} \Phi \overline{(N_R^\alpha)^c} N_R^\alpha + {\rm H.c.,} 
\label{eq:LYk}   
\eea
where we start-off in a basis where the $Y_{N_{\alpha}}$ matrix is diagonal, and $\tilde{H} = i \tau^2 H^*$ with $\tau^2$ being the second Pauli matrix. The scalar potential involving two scalar fields is given by 
\bea
V=\sum_{\mathcal{I}= H, \Phi} \Big[m_{\mathcal{I}}^2 (\mathcal{I}^{\dagger} \mathcal{I})+ \lambda_{\mathcal{I}} (\mathcal{I}^\dagger \mathcal{I})^2 \Big] +
\lambda_{\rm mix} (H^\dagger H) (\Phi^{\dagger} \Phi)\,.
\label{pot}
\eea 

After the breaking of B$-$L and electroweak gauge symmetries, the scalar fields $H$ and $\Phi$ develop their VEVs as 
\begin{align}\label{eq:VEV}
  \langle H\rangle \ = \ \frac{1}{\sqrt{2}}\begin{pmatrix} v+h\\0 
  \end{pmatrix}~, \quad {\rm and}\quad 
  \langle\Phi\rangle \ =\  \frac{v_\Phi^{}+\phi}{\sqrt{2}}~,
\end{align}
where electroweak scale is  $v=246$ GeV at the potential minimum. For $v_\Phi\gg v$, the mass of the B$-$L gauge boson can be written as $M_{Z^\prime}^{}= 2 g_X  v_\Phi$. The breaking of B$-$L symmetry induces the Majorana mass term for the RHNs while the electroweak symmetry breaking generates the Dirac mass term for the light left-handed neutrinos from Eq.~\eqref{eq:LYk} as
\begin{equation}
    M_{\alpha}^{} \ = \ \frac{Y_{N_\alpha}}{\sqrt{2}} v_\Phi^{}, \, \, \, \, \,
    m_{{D}_{\alpha \beta}} \  =  \ \frac{Y_{{\nu}_{\alpha \beta}}}{\sqrt{2}} v\,.
\label{eq:mDI}
\end{equation}
From the mass matrices above, the light active neutrino masses can be derived using the standard see-saw formula $-m_D^{} M_\alpha^{-1} m_D^T$~\cite{Gell-Mann:1979vob, Sawada:1979dis, Mohapatra:1980yp}. This mechanism successfully explains the tiny neutrino masses and their flavor mixing. 
%%%%%%%%%%%%%%%%
\begin{figure*}[htb!]
\centering        
\includegraphics[scale=.37]{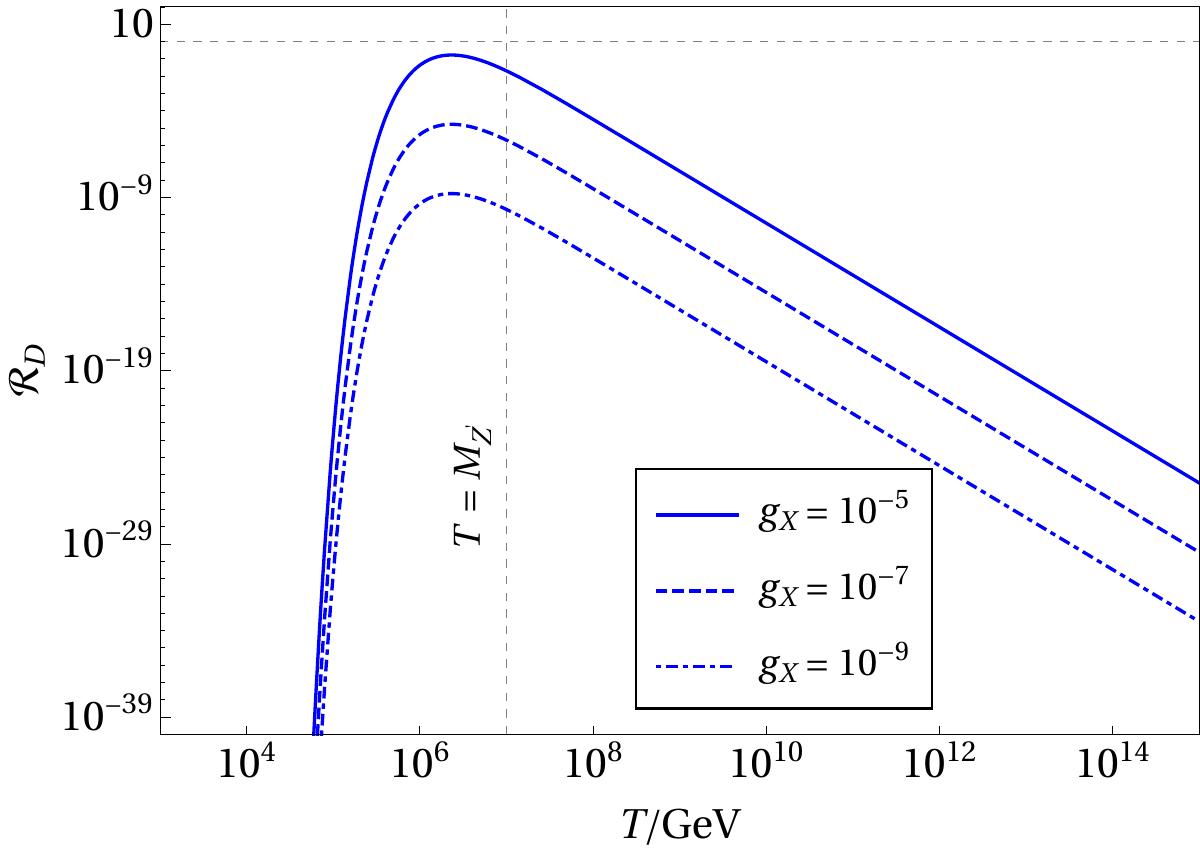}~\includegraphics[scale=.37]{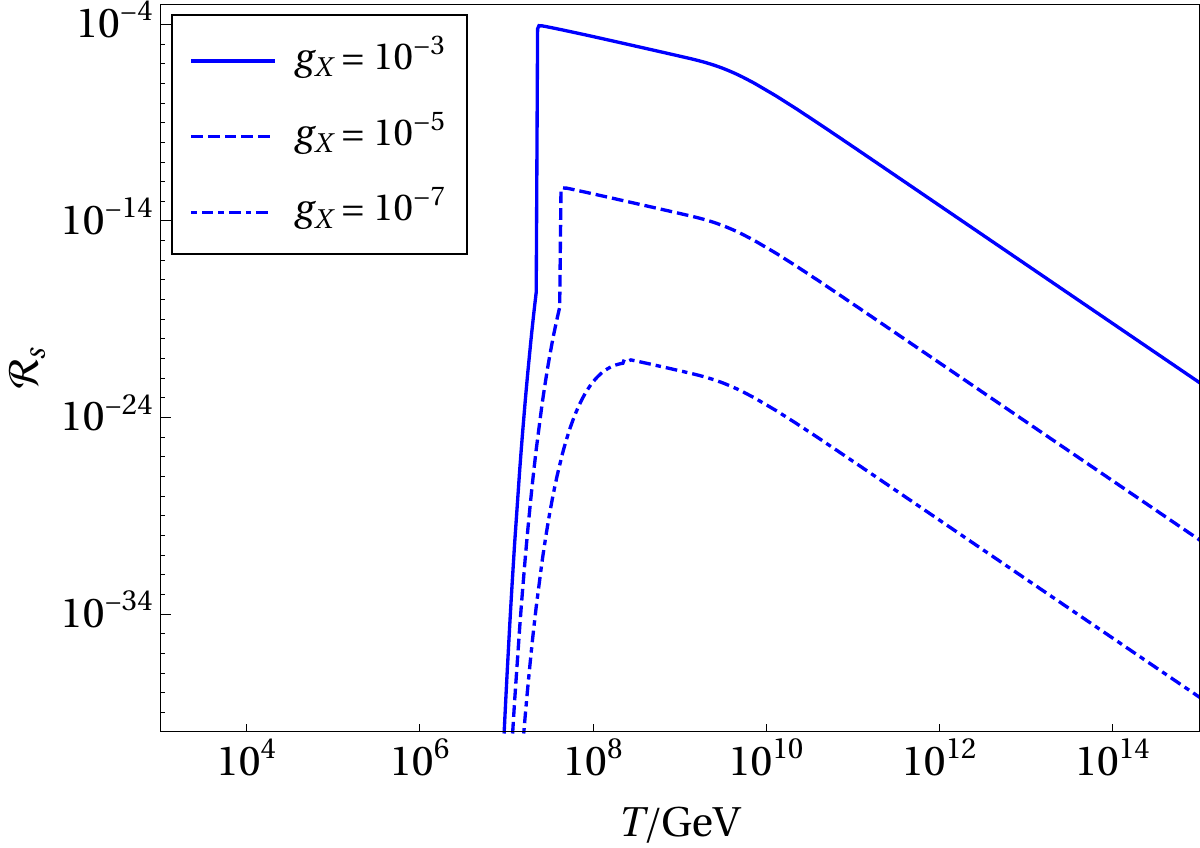}
\caption{{\it Left:} DM production rate from $Z^\prime$ decay, as a function of the bath temperature $T$, for different choices of $g_X$. Here we have fixed $\Mzp=10^7$ GeV. {\it Right:} Same as left, but considering DM production from 2-to-2 scattering, mediated by $Z^\prime$, where $\Mzp=10^5$ GeV. In all cases the DM mass is fixed at 4 PeV.}
\label{fig:rate}
\end{figure*}
%%%%%%%%%%%%%%%%%%%%%%%%%%
\begin{figure*}[htb!]
\centering        
\includegraphics[scale=.37]{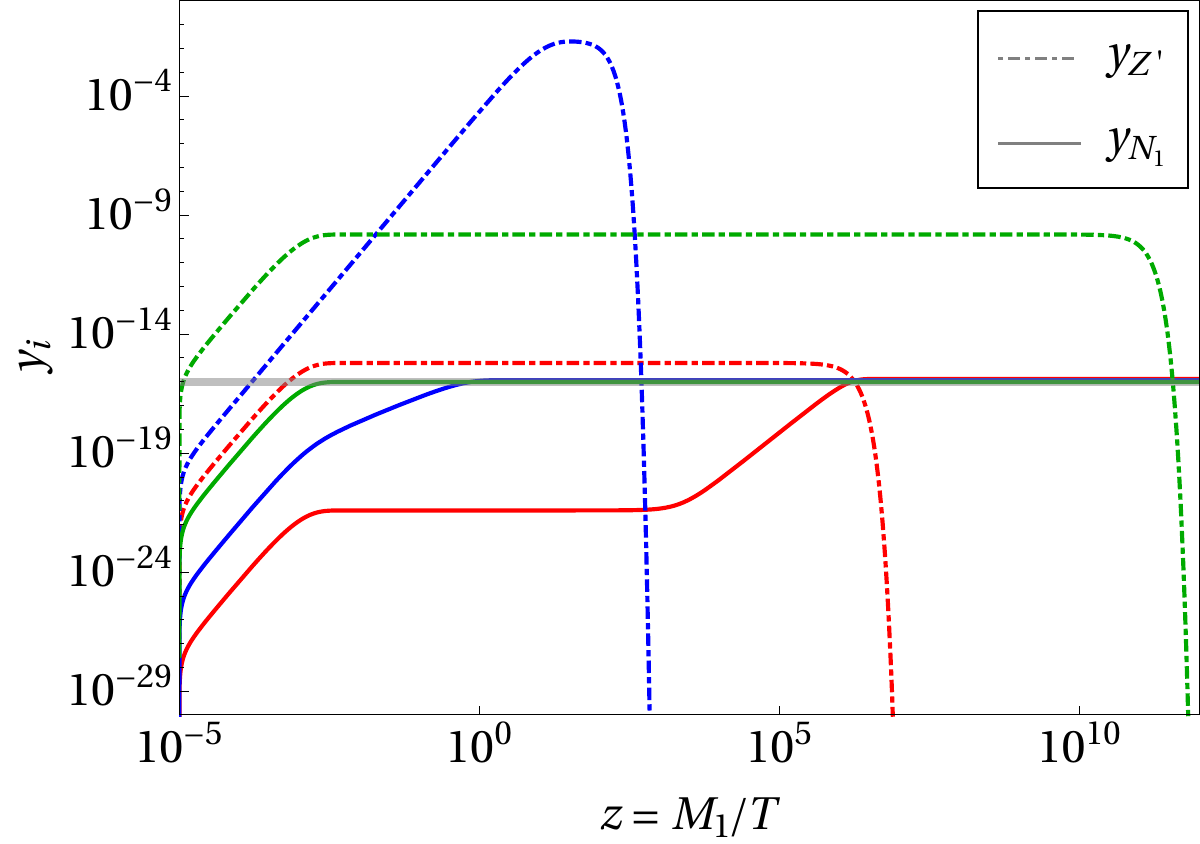}~\includegraphics[scale=.37]{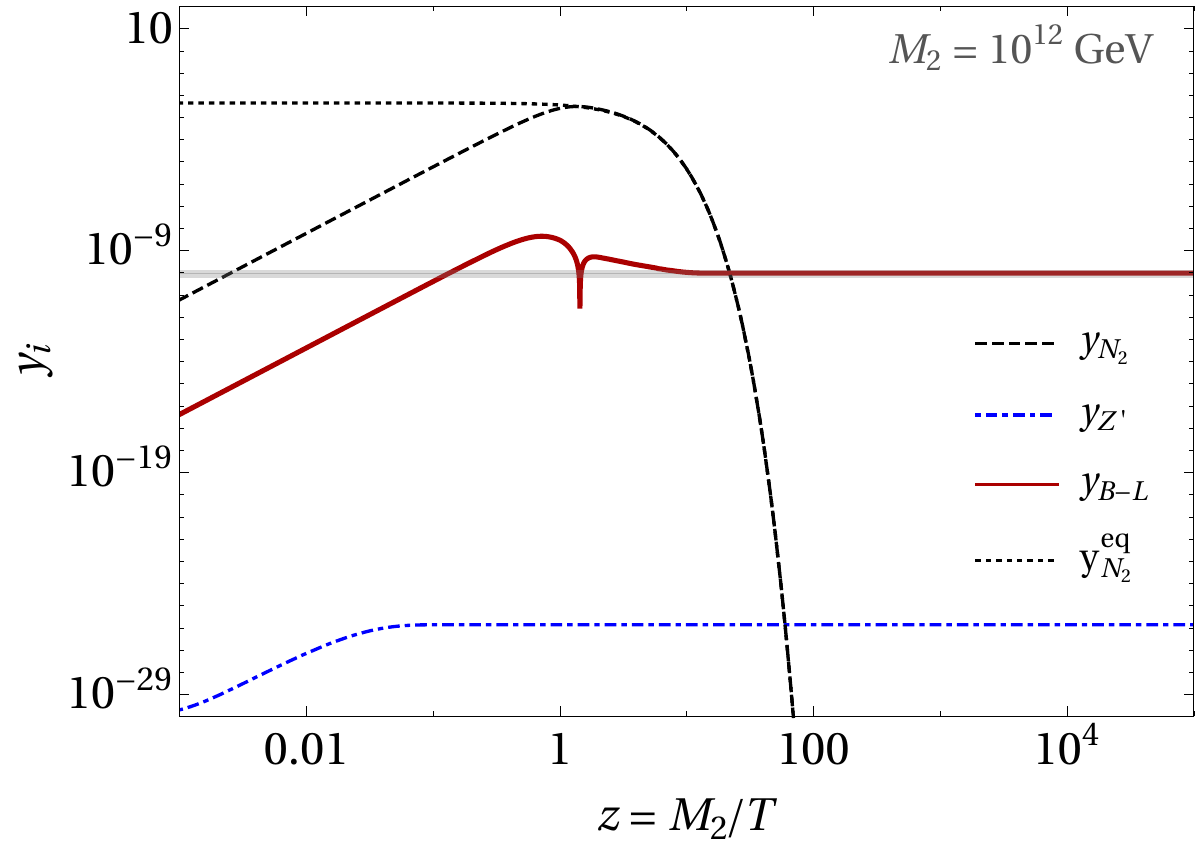}
\caption{In all cases DM mass if fixed to $M_{N_1}=4$ PeV. {\it Left:}  Evolution of DM and $Z'$ yields in solid and dashed curves, respectively, as function of $z=M_1/T$. The red, blue and green colours correspond to $\Mzp=\{10^8,\,10^5,\,10^3\}$ GeV with the corresponding $g_X=\{2\times 10^{-12},\,1.2\times 10^{-6},\, 10^{-14}\}$, respectively, to produce the right DM abundance (shown via gray thick straight line). For the red and green curves we fix $m_\phi=10^{10}$ GeV, while for the blue curve $m_\phi=10^6$ GeV. {\it Right:} Evolution of B$-$L yield (red solid) with $M_2/T$, for $N_2$ leptogenesis. The dark red curve produces the observed baryon asymmetry (shown via gray thick straight line). We show for comparison the equilibrium $N_2$ yield $y_{N_2}^{\rm eq}$ in black dotted curve. }
\label{fig:yld}
\end{figure*}
%%%%%%%%%%%%%%%%%%%%%
Out of three generations of RHNs, we identify $N_1$ as a long-lived decaying DM. Its only decay channel, $N_1\to \ell\,H$, produces boosted high-energy neutrinos. The DM lifetime, as required by the IceCube observations, in terms of its mass and Yukawa coupling reads~\cite{Esmaili:2014rma,Higaki:2014dwa,Chianese:2016smc,Arguelles:2022nbl}  
\bea\label{eq:life-N1}
\tau\simeq10^{28}~\text{s}\,\left(\frac{\left(Y_{\nu}\right)_{11}}{2\times 10^{-29}}\right)^2\,\left(\frac{M_1}{4\,\text{PeV}}\right)\,. 
\eea
Due to extremely small Yukawa coupling strength, it is not possible to address right DM abundance via the inverse decay channel $\ell\,H\to N_1$. Thus, we consider its production via freeze-in~\cite{Hall:2009bx,Bernal:2017kxu} from the thermal bath. Consequently, production and decay of $N_1$ is disentangled, and hence right DM abundance can be satisfied even with a DM decay lifetime of $\tau_{N_1}\sim 10^{28}$ sec, complying with the IceCube data. Here onward we will consider the following set of parameters as independent in our analysis: $\{g_X,\,\Mzp,\,Y_{N_{\alpha}}\}$, while the DM mass ($M_1 = Y_{N_1}v_{\Phi}/\sqrt{2}$) is always fixed at 4 PeV. Note that, the long-lived RHN $N_1$, which is a potential DM candidate, contributes negligibly to the active neutrino masses because of the tiny Yukawa coupling $\left(Y_{\nu}\right)_{11}\simeq\mathcal{O}(10^{-30})$ [cf. Eq.~\eqref{eq:life-N1}]. As a result, the lightest active neutrino remains (almost) massless. This scenario is fully compatible with current neutrino oscillation data, which constrain only the mass-squared differences and not the absolute mass scale. This also provides a complementary prospect in experiments sensitive to the absolute neutrino mass scale. While such a tiny neutrino mass is out of reach of ongoing tritium beta decay experiments like KATRIN~\cite{KATRIN:2019yun}, future observation of neutrinoless double beta decay can falsify our scenario~\cite{Dolinski:2019nrj}, particularly for normal ordering.
\\

%%%%%%%%%%%%%%%%%%%%
\noindent
{\textbf{DM production via freeze-in}--}
%%%%%%%%%%%%%%%%%%%
The PeV scale $N_1$ DM can be produced via (i) on-shell decay of $Z^\prime$, provided $M_{Z^{\prime}}> 2\,M_1$, (ii) on-shell decay of $\phi$, if $m_\phi> 2\,M_1$ and (iii) 2-to-2 scattering off of the bath particles, mediated by $Z^\prime$. Notice that we consider the limit of zero mixing\footnote{Current bounds from LHC~\cite{Robens:2015gla,Chalons:2016jeu,Das:2022oyx}, LEP~\cite{LEPWorkingGroupforHiggsbosonsearches:2003ing} prospective colliders like ILC~\cite{Wang:2020lkq} and CLIC~\cite{CLIC:2018fvx} suggests that the scalar mixing requires to be $<0.001$, for $\phi$ mass up to 1 TeV. Since in our case $m_\phi\gg 1$ TeV, therefore we consider the mixing to be vanishingly small.} between the extra scalar and the SM Higgs thus we do not include any $\phi$ mediated scattering in our analysis\footnote{Some dedicated studies of scalar effects on the  DM production via freeze-in include \cite{Kaneta:2016vkq, Eijima:2022dec, Seto:2024lik}.}. The universal interaction strength $g_X$ must be feeble to ensure non-thermal DM production via freeze-in. As a result, the $Z^\prime$ cannot reach thermal equilibrium, and its comoving number density must be determined by solving a set of coupled Boltzmann equations (BEQs) are given in Appendix~\ref{sec:BEQs}. The first line of Eq.~\eqref{eq:cBEQ} corresponds to the yield of $\phi$, with $\Gamma_\phi$ being its decay rate as reported in Appendix.~\ref{sec:decays}. The second line of Eq.~\eqref{eq:cBEQ} takes care of the $Z^\prime$ yield, sourced from the decay of $\phi\to Z'\,Z'$. Finally, in the third line, we provide the evolution equation for DM yield, sourced from $Z^\prime$-decay, $\phi$-decay and $Z^\prime$-mediated scattering channels. Since the mixing between $\phi$ and the SM Higgs is considered to be negligibly small, its decay into SM final states are extremely suppressed. Consequently $\phi$ decays dominantly into RHNs and $Z^\prime$ pairs. 
%%%%%%%%%%%%%%%%%%%%%%%%
\begin{figure*}[htb!]
\centering        
\includegraphics[scale=.45]{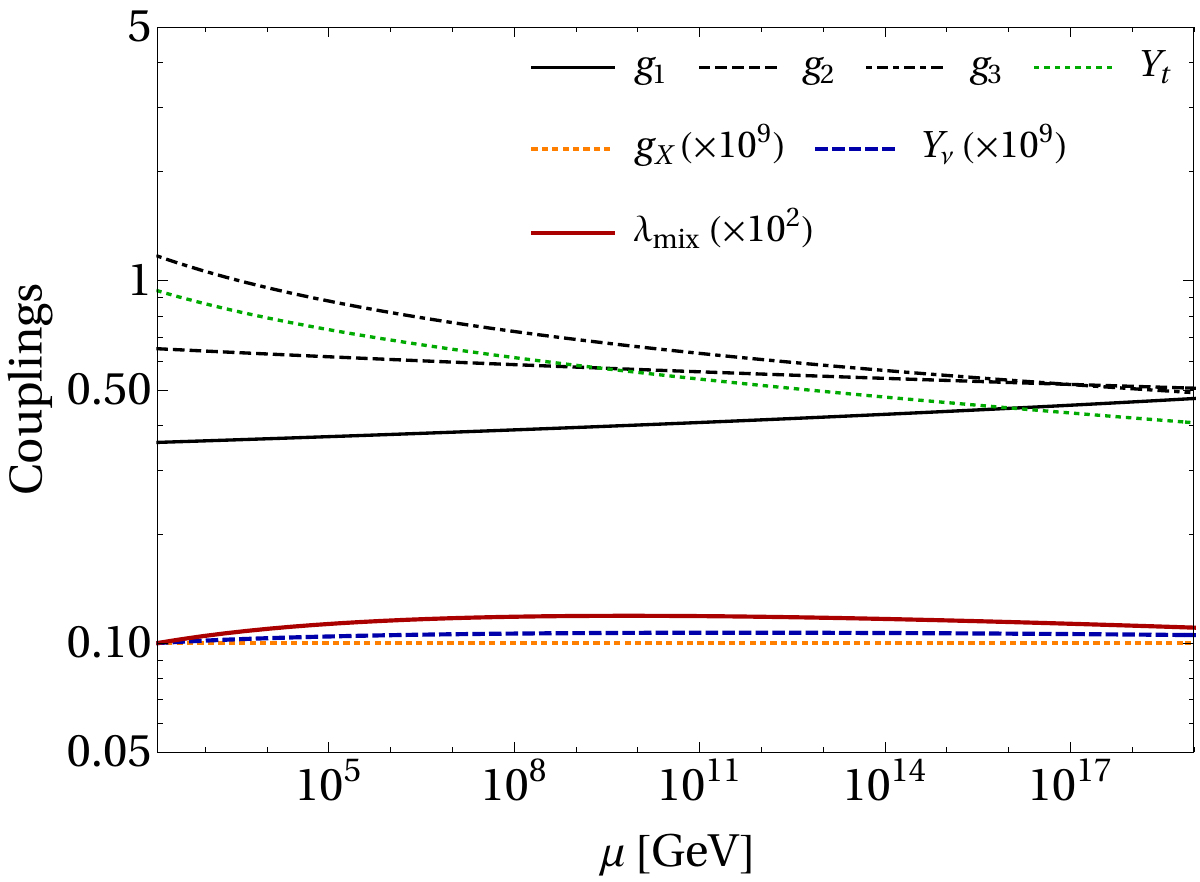}
\caption{Evolution of different couplings appearing in the present framework with the running scale $\mu:\left[m_t,\,M_P\right]$, with the top quark mass $m_t=173.2$ GeV. Note that, $g_X\,\text{(shown in orange dot)},\,Y_\nu\,\text{(shown in blue dashed)},\,\lambda_{\rm mix} \,\text{(shown in red solid)}$ are magnified.}
\label{fig:rge}
\end{figure*}
%%%%%%%%%%%%%%%%%%
To fit the observed DM relic density, it is required that $Y_0\, M_1 = \Omega h^2 \, \frac{1}{s_0}\,\frac{\rho_c}{h^2} \simeq 4.3 \times 10^{-10}\,\text{ GeV}$,
where $Y_0 \equiv y_{N_1}(z\to\infty)$ is the present DM yield. We use the critical energy density $\rho_c \simeq 1.05 \times 10^{-5}\, h^2$~GeV/cm$^3$, present entropy density $s_0\simeq 2.69 \times 10^3$~cm$^{-3}$~\cite{ParticleDataGroup:2022pth} and DM relic abundance $\Omega h^2 \simeq 0.12$, with $h\simeq H_0/100\,\left({\rm km/s/Mpc}\right)$ being the reduced Hubble rate, where $H_0\simeq 67.4 \pm 0.5 \text{ km/s/Mpc}$ is the current Hubble rate~\cite{Planck:2018vyg}.
%%%%%%%%%%%%%%%%%%%%%%%%%%%

Before moving on, let us emphasize that for the freeze-in to be valid, the DM has to be
out of chemical equilibrium with the SM bath.
Now, the DM production rate from $Z^\prime$ decay  is given by the production rate densities as
\begin{align}
\gamma_D=\frac{g_a}{2\,\pi^2}\,m_a^2\,\Gamma_{Z^\prime\to N_1\,N_1}\,T\,K_1\left(\frac{m_a}{T}\right)\,,
\end{align}
while production from 2-to-2 scattering off of the bath particles reads
\begin{align}
& \gamma_s=\frac{T}{32\pi^4}\,g_a g_b\times\int_{\text{max}\left[\left(m_a+m_b\right)^2,4M_1^2\right]}^\infty ds
\nonumber\\&
\frac{\biggl[\bigl(s-m_a^2-m_b^2\bigr)^2-4m_a^2 m_b^2\biggr]}{\sqrt{s}}\,\sigma\left(s\right)_{a,b\to N_1\,N_1}\,K_1\left(\frac{\sqrt{s}}{T}\right)\label{eq:gam-ann}\,. 
\end{align}
Here, the subindices $a,b$ correspond to the SM states, $g_{a,b}$ are the corresponding degrees of freedom and $K_i$ denotes the modified Bessel functions of $i^{\rm th}$ kind. In order to ensure out of equilibrium DM production, the following condition needs to be satisfied
\begin{align}
& \mathcal{R}_{D(s)}\equiv\frac{\gamma_{D(s)}}{n_{\rm eq}^{N_1}\,H} < 1\,,    
\end{align}
where $H$ is the Hubble parameter during radiation domination and $n_{\rm eq}=(T/(2\pi^2))\,m^2\,K_2\left(m/T\right)$ is the equilibrium number density. Note that, here we use the equilibrium number density for the DM itself in order to obtain the conservative bound on the masses and couplings. In Fig.~\ref{fig:rate} we show the evolution of DM production rate, as a function of the temperature of the thermal bath. Due to $g_X^2$ dependence, it is not possible to have $g_X\gtrsim 10^{-4}$ in order to ensure $\mathcal{R}<1$ till $T\simeq\Mzp$ in the case of DM genesis via decay. However, for scattering it is still possible to have $g_X$ as large as about $10^{-3}$, while ensuring $\mathcal{R}<1$ because of the $g_X^4$ dependence of the scattering cross-section. Note that, in this case the DM production starts at $T\sim M_1$, as prior to that the thermal bath is not energetic enough to produce such a massive DM.
%%%%%%%%%%%%%%%%%%%%%%%%%%%%%%%%%%%
\begin{figure*}[htb!]
    \centering       
    \includegraphics[scale=.55]{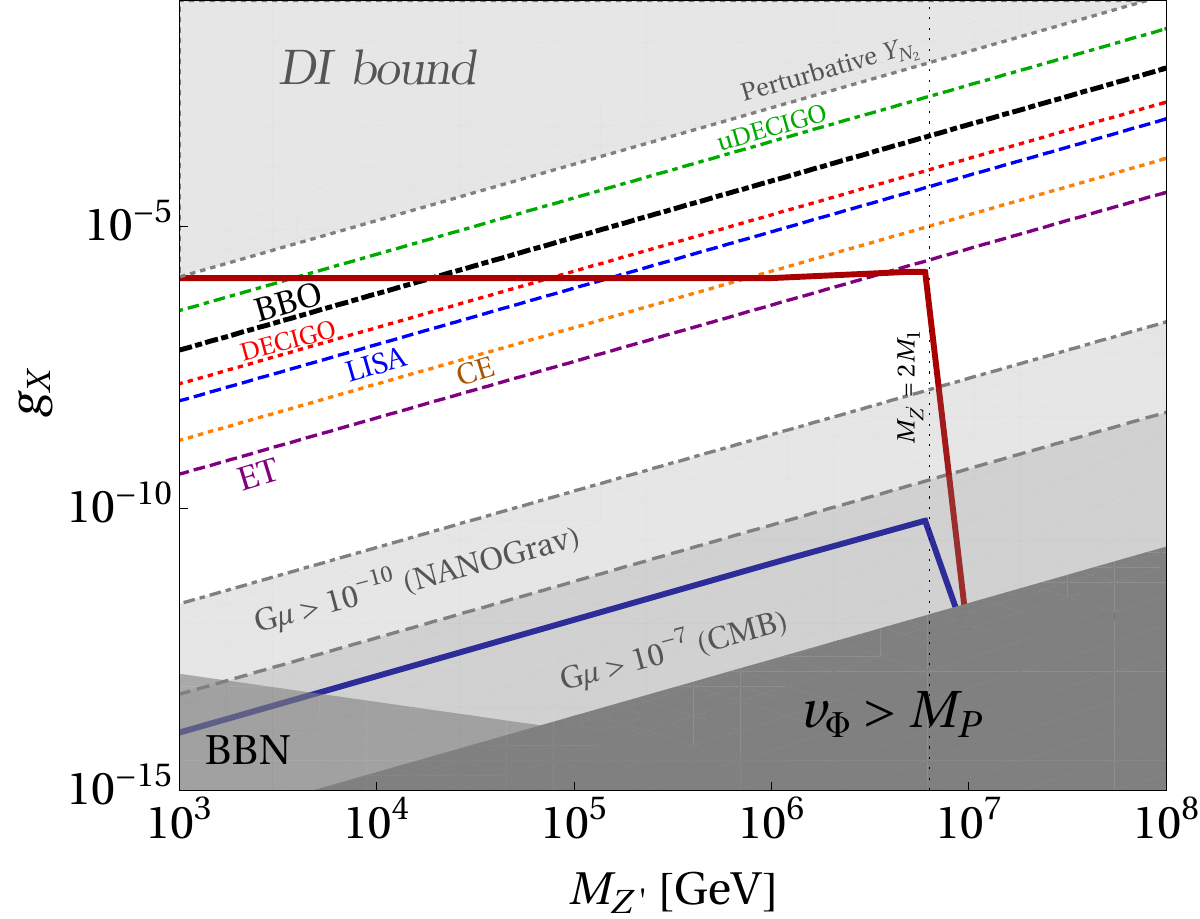} 
    \caption{{\it Summary of parameter space:} The red and blue thick contours correspond to right DM abundance by fixing $M_1=4$ PeV. Along the  red contour we consider DM production both via scattering and decay, while along the blue contour, only production via decay is considered (see text).  The diagonal contours correspond to different $v_\phi$'s, some of which fall within the sensitivity of a few GW detectors, as denoted. The shaded regions are disallowed from Davidson-Ibarra (DI) bound $(M_2<10^9\,\text{GeV})$, BBN bound on $Z^\prime$-lifetime and by having super-Planckian VEV-values.}
    \label{fig:paramspace}
\end{figure*}
%%%%%%%%%%%%%%%%%%%%%%%%%%%%%%%%%%%%%%%%%%%

The evolution of DM and $Z^\prime$ yields $(y_{Z^\prime})$ are shown in the left panel of Fig.~\ref{fig:yld}, as a function of $z=M_1/T$, with $M_1=4$ PeV. For the red curves, we fixed $m_\phi/2>\Mzp>2\,M_{N_1}$, such that the DM can be produced from the on-shell decay of $Z^\prime$, as well as via $Z^\prime$ mediated scattering. Now, in presence of both  channels, DM production via decay dominates since the production rate is proportional to $g_X^2$, compared to scattering, which is proportional to $g_X^4$. The DM yield $y_{N_1}$ gradually grows with time, finally reaching the observed abundance in the asymptotic limit. In the first plateau region of the red solid curve, DM is dominantly produced from $\phi$-decay. Due to large decay width, $\phi$ decay is completed earlier and the DM (as well as $Z^\prime$) production from $\phi$-decay stops. However, $Z^\prime$ decay still goes on. Thus, the final DM abundance is set by $Z^\prime\to N_1\,N_1$ channel. On the other hand, the blue solid curve represents a scenario where the DM production is instead controlled by $Z'$ mediated scattering, as the decay is kinematically forbidden when $\Mzp<M_1$, also $m_\phi<2\,M_1$, but $m_\phi>2\,\Mzp$ such that $Z'$'s are produced from on-shell $\phi$-decay. For $M_{Z'}=10^8\,(10^5)$ GeV and for DM mass of 4 PeV, the freeze-in occurs at the temperature $T\sim \Mzp\,(M_1)$. This is the typical IR feature~\cite{Bernal:2017kxu}, where the DM yield saturates when $T={\rm max}\left[\Mzp,\,M_1\right]$. Finally, for $\Mzp/2<M_1\ll m_\phi$, the DM is dominantly produced from $\phi$-decay (green solid curve), due to which, once again, a very tiny $g_X$ is required since this decay rate is also proportional to $g_X^2$.
\\

%%%%%%%%%%%%%%%%%%%%
\noindent
{\textbf{Baryon asymmetry from $N_2$ leptogenesis}--}
%%%%%%%%%%%%%%%%%%%
The observed BAU is generated via vanilla leptogenesis from the CP-violating, out of equilibrium decays of $N_2$. Due to very long lifetime, the lepton asymmetry generated through $N_1$ decay does not contribute to the generation of the BAU as its decay takes place far below the sphaleron equilibration temperature. The CP asymmetry from $N_2$-decay can be expressed as~\cite{Buchmuller:2004nz,Kaneta:2019yjn} 
\begin{align}\label{eq:epsL}
& \epsilon_{\Delta L}\simeq \frac{3 \delta_{\rm eff}}{16\,\pi}\,\frac{M_2\,m_{\nu\,,\text{max}}}{v^2}\,, 
\end{align}
where $\delta_{\text{eff}}$ is the effective CP violating phase in the neutrino mass matrix with $0\leq \delta_{\rm eff}\leq 1$ (see Appendix.~\ref{sec:CP}) and we take $m_{\nu,\text{max}} = 0.05$ eV as the heaviest light neutrino mass following the normal hierarchy. The final B$-$L asymmetry is obtained by solving a set of coupled BEQs given in the Appendix~\ref{sec:N2lepto}. Once again, the first line of the set of equations above takes into account the production of $Z^\prime$ from the thermal bath, via the decay of $\phi$ and the decay of $Z^\prime$ into several final states. The second line corresponds to the evolution of $N_2$ yield. The produced $N_2$'s then undergo CP violating out of equilibrium decay producing a net B$-$L asymmetry. The evolution of this asymmetry is tracked by the last line of the equation, where we also include the  contribution of inverse decays to the washout term.  

The sphaleron interactions~\cite{Buchmuller:2004nz} are in equilibrium within the temperature range 100 GeV to $10^{12}$ GeV, and they convert a fraction of a non-zero B$-$L asymmetry into a baryon asymmetry via
\begin{align}
Y_B\simeq a_{\rm sph}\,y_{B-L}=\frac{8\,N_F+4\,N_H}{22\,N_F+13\,N_H}\,y_{B-L}\,,    
\end{align}
where $N_F(=3)$ and $N_H(=1)$ are numbers of fermion generations and Higgs doublets, $a_{\rm sph}\simeq 28/79$ and $Y_B^o \simeq 8.75\times 10^{-11}$. We show the evolution of yields for RHN $(y_{N_2})$ and B$-$L $(y_{\rm{B-L}})$ as a function of $z=M_2/T$ in the right panel of Fig.~\ref{fig:yld}. The population of $N_2$ receives contributions from two sources: (i) the thermal bath, through the inverse decay process $\ell\,H \to N_2$ and (ii) decay of $Z^\prime$ and $\phi$. Notably, the bath contribution is the dominant one, as all other production channels are suppressed due to tiny $g_X$ to satisfy freeze-in conditions. Following the Davidson-Ibarra (DI) bound~\cite{Davidson:2002qv}, $M_2\gtrsim 10^9$ GeV is required for successful thermal leptogenesis. Since the RHN mass depends on $v_\Phi$ from Eq.~\eqref{eq:mDI}, in order to satisfy DI bound, $v_\Phi\gtrsim 10^9$ GeV for $Y_{N_2}\sim\mathcal{O}(1)$.
\\

\noindent
{\textbf{Running of the couplings}--}
To solve the renormalization group equations (RGEs) and track the running of the couplings, we set initial conditions at the top quark mass scale $\mu = m_t = 172.4~\mathrm{GeV}$. The SM parameters, including the top Yukawa coupling $Y_t$, are taken from~\cite{Buttazzo:2013uya}, while we fix the new physics couplings as $g_X = Y_\nu = 10^{-10}$. For the scalar quartic couplings, we choose $\lambda_H = 0.12$, $\lambda_\Phi = 10^{-2}$, and $\lambda_{\rm mix} = 10^{-3}$. With these initial values, we evolve the couplings up to the Planck scale, $\mu = M_P$, considering the set of equations in Appendix.~\ref{sec:RGE}. Given that $\beta_X \sim g_X^3$, the new gauge coupling $g_X$ remains nearly constant due to its tiny initial value, as shown by the orange dashed curve in Fig.~\ref{fig:rge}. The same applies to the Yukawa coupling $Y_\nu$, illustrated by the blue dashed curve. The scalar mixing coupling $\lambda_{\rm mix}$ also stays approximately unchanged (red solid curve), primarily because of its small initial value constrained by collider bounds. Its slight downward bend near $\mu \to M_P$ arises from the negative contribution of the term $-\frac{9}{2}\,g_2^2\,\lambda_{\rm mix}$ in its $\beta$-function.
\\

\noindent
{\textbf{GW Spectrum from cosmic strings}--} Numerical simulations based on the Nambu-Goto action \cite{Ringeval:2005kr,Blanco-Pillado:2011egf} have found that for gauged symmetry, the dominant channel of energy loss from CS is through GW radiation from oscillating loops\footnote{The present framework might also give rise to primordial GW signal via graviton Bremsstrahlung through decay~\cite{Nakayama:2018ptw,Barman:2023ymn,Barman:2023rpg,Choi:2024acs}, considering {\it minimal} gravitational interactions. In the context of leptogenesis this has been explored in~\cite{Ghoshal:2022kqp,Datta:2024tne}. However, such decays are suppressed by the Planck mass. The amplitude of Bremsstrahlung GW signal is also suppressed for the same reason, and their spectral peak typically lies in the GHz range\---accessible only to specialized detectors, such as resonant cavities. If one introduces a non-minimal coupling between the DM and the Ricci scalar, then the DM can also decay via the non-minimal gravity portal~\cite{Cata:2016epa,Cata:2017jar}, however we do not introduce such coupling in this analysis.}. The rate of energy loss or the power of GW emission is given by~\cite{Vilenkin:1981bx},
$P_{\rm GW} = \frac{G}{5} (\dddot{Q})^2 $, where $Q$ is the quadrupole moment of the oscillating loop, and the triple time derivative $\dddot{Q}\propto \mu$. Thus, the rate of energy loss can be written as
$\frac{dE}{dt}=-\Gamma G \mu^2$, where $\Gamma\approx 50$~\cite{Vachaspati:1984gt}. Due to GW emission, the loop starts to shrink from its initial length $l_i = \alpha t_i$ at the time of formation $(t_i)$ as $l(t)= \alpha t_i - \Gamma G \mu (t-t_i)$, where $\alpha$ is the loop size parameter and considered to be $\alpha \approx 0.1$ from simulation studies \cite{Blanco-Pillado:2013qja,Blanco-Pillado:2017oxo}. The total energy loss from a loop constitutes a set of normal mode oscillations with frequencies $f_k=2k/l(t)$, where  $k$ represents number of modes $(k=1,2,3...\infty)$. The GW density parameter is defined as
\begin{align}
    \Omega_{\rm GW}(t_0,f)=\frac{f}{\rho_c}\frac{d\rho_{\rm GW}(t_0,f)}{df}=\sum_k\Omega_{\rm GW}^{(k)}(t_0,f)\,,\label{eqn:omgcs1}
\end{align}
where $f$ and $t_0$ represent the current frequency and the present time, while $\rho_c= 3 M_P^2 H_0^2$ is the critical energy density where $M_P$ is reduced Planck mass and Hubble parameter. Since the GW energy density redshifts as $a^{-4}$, we have \cite{Blanco-Pillado:2013qja}
\begin{align}
    \frac{d\rho_{\rm GW}^{(k)}}{df}=\int_{t_F}^{t_0} \left[\frac{a(t_E)}{a(t_0)}\right]^4 P_{\rm GW}(t_E,f_k)\frac{dF}{df}dt_E\,,\label{eqn:omgcs2}
\end{align}
where $f_k$ denotes emitted frequency $(f_E)$ at the time $t_E$, $t_F$ is the loop formation time, $\frac{dF}{df}=f \left[\frac{a(t_0)}{a(t_E)}\right]$ accounts for the redshift of the frequency, and 
\begin{align}
    P_{\rm GW}(t_E,f_k)     =\frac{2\,k\,G\mu^2\,\Gamma_k}{f\left[\frac{a(t_0)}{a(t_E)}\right]^2}\,n\left(t_E,\frac{2k}{f}\,\left[\frac{a(t_E)}{a(t_0)}\right]\right)\,,
     \label{eqn:omgcs3}
\end{align}
is the power emitted by the loops. 

The GW spectrum depends on the nature of small-scale structure in the loops which can appear in the form of cusps or kinks~\cite{Damour:2001bk,Gouttenoire:2019kij}. For our study, we consider cusp-like structures to dominate the GW spectra. Here we have $\Gamma_k=\frac{\Gamma k^{-4/3}}{\sum_{m=1}^\infty m^{-4/3}}$, with $\sum_k\Gamma_k=\Gamma$ and $n$ denotes the number density of loops which, in a cosmological background with scale factor $a\propto t^{\beta}$, is found from the Velocity dependent One Scale (VOS) model~\cite{Martins:1996jp,Martins:2000cs,Auclair:2019wcv} and numerical simulations to be \cite{Blanco-Pillado:2013qja}
\begin{align}
    n(t_E,l_{k}(t_E))=\frac{A_\beta}{\alpha}\frac{(\alpha+\Gamma G \mu)^{3(1-\beta)}}{\left[l_k(t_E)+\Gamma G \mu t_E\right]^{4-3\beta}t_E^{3\beta}}\,,
    \label{eqn:omgcs4}
\end{align}
where $A_\beta$ is a constant depending on the cosmological background. Using Eqs.~\eqref{eqn:omgcs1}-\eqref{eqn:omgcs4}, we obtain the current GW energy density for the mode $k$ as
\begin{align}
    \Omega_{\rm GW}^{(k)}(t_0,f)=\frac{2k G\mu^2\Gamma_k}{f \rho_c}\int_{t_{osc}}^{t_0}dt\left[\frac{a(t)}{a(t_0)}\right]^5 n\left(t,l_k\right)\,,
    \label{eqn:omgcsfin}
\end{align}
where the integration over time is from the moment $t_{\rm osc}$ when loops start oscillating after damping due to thermal friction~\cite{Vilenkin:1991zk} and hence producing sub-dominant effect. For the case of loops formed and radiated during radiation domination, the GW spectrum has a typical flat plateau, with the amplitude given by 
\bea
\Omega_{\rm GW}^{(k=1),{\rm plateau}}(f)=\frac{128\pi G\mu}{9\zeta(4/3)}\frac{A_r}{\epsilon_r}\Omega_r\left[(1+\epsilon_r)^{3/2}-1\right]\,,
\eea    
where $\epsilon_r=\alpha/\Gamma G\mu$, and $A_r = 0.54$~\cite{Auclair:2019wcv} for radiation domination. CMB measurements require $G\mu\lesssim 10^{-7}$~\cite{Charnock:2016nzm} and hence we have $\alpha \gg \Gamma G \mu$. This gives $\Omega_{\rm GW}^{(k=1)}(f)\propto  v_{\Phi}$, and hence a higher symmetry breaking scale is more likely to be probed by the GW detectors. Recent results from NANOGrav~\cite{NANOGrav:2023hvm} puts a strong upper bound on  $G\mu\lesssim 10^{-10}$. However, this bound is found to be relaxed, considering long-lived loops in a wider class of models~\cite{Kume:2024adn}, with the upper bound close to the CMB limit $G\mu\lesssim10^{-7}$ \cite{Charnock:2016nzm, Lizarraga:2016onn}.
%%%%%%%%%%%%%%%%%%%%%%%%

The viable parameter space is furnished in Fig.~\ref{fig:paramspace}, where the thick-solid red and blue contours correspond to the observed DM abundance for $M_{N_1}=4$ PeV. The diagonal dashed lines denote different $v_{\Phi}$ providing the correct value of $\Mzp$ for the corresponding $g_X$. A few of such $v_\Phi$'s are within the reach of proposed GW detectors: Big Bang Observer (BBO)~\cite{Crowder:2005nr, Corbin:2005ny}, ultimate DECIGO (uDECIGO)~\cite{Seto:2001qf, Kudoh:2005as}, LISA~\cite{LISA:2017pwj}, the cosmic explorer (CE)~\cite{Reitze:2019iox} and the Einstein Telescope (ET)~\cite{Hild:2010id, Punturo:2010zz, Sathyaprakash:2012jk, Maggiore:2019uih}. The gray shaded region in top left corner is disallowed by the DI bound on the RHN mass for $Y_{N_2}\lesssim\sqrt{4\pi}$. The light gray shaded region in the bottom left corner corresponds to the combination of $g_X$ and $\Mzp$ that gives rise to $\tau_{Z^\prime}=1/\Gamma_{Z^\prime}>1$ sec, thereby jeopardizing the big bang nucleosynthesis (BBN) predictions. The dark gray shaded region in the bottom right corner demands (super-)Planckian $v_\Phi$. As the $Z^\prime\to N_1\,N_1$ decay channel opens up, one needs $g_X\lesssim 10^{-10}$ to satisfy the non-thermal  freeze-in condition in order to produce the correct relic abundance. As a result, the thick-solid red and blue lines take a sharp bend at the mass threshold. Along the horizontal part of the red contour, we choose $m_\phi=10^6$ GeV, such that DM is produced solely from $Z'$-mediated scattering, while $Z'$ is produced from $\phi$ decay. In this case, the final DM yield approximately reads $Y(T_{\rm FI})\sim\sigma(T_{\rm FI})\,M_P\,T_{\rm FI}\sim g_X^4\,M_P/M_1$, where $\sigma(T_{\rm FI})\sim g_X^4/T_{\rm FI}^2$, with freeze-in temperature $T_{\rm FI}\sim M_1$. Consequently, the DM abundance becomes almost independent of the mediator mass. For the thick-solid blue contour, $m_\phi=10^{10}\,\text{GeV}>2\,M_1$, as well as $m_\phi>2\,\Mzp$ (and $\Mzp>2\,M_1$). As a result, the DM production is decay-dominated, demanding $g_X$ to be extremely small, following Fig.~\ref{fig:yld}.
%%%%%%%%%%%%%%%%%%%%%%%%
\\

\noindent
{\textbf{Conclusions}--} We have proposed a scenario that simultaneously accounts for both DM and the BAU while remaining testable at IceCube as well as at several proposed GW detectors. Notably, our approach is broadly applicable to the general class of $U(1)_X$ models. A key element of our framework is the symmetry breaking scale, which unifies all BSM masses and couplings, offering a coherent and compact picture, as illustrated in Fig.~\ref{fig:paramspace}. This provides bounds on extremely small $g_X$ along with $\Mzp\gtrsim$ 1 TeV, both lying beyond the reach of standard collider experiments. This work underscores the essential role of multi-messenger astronomy in probing feebly coupled new physics.
\\

%%%%%%%%%%%%%%%%%%%%%
\noindent
{\textbf{Acknowledgments}--}The work of SJD was supported by IBS under the project code, IBS-R018- D1. We thank Osamu Seto and Rinku Maji for useful communications.
%%%%%%%%%%
\begin{widetext}
\appendix
%%%%%%%%%%%%%%%
\section{Relevant decay widths and cross-sections}
\label{sec:decays}
%%%%%%%%%%%%%
The partial decay width of $Z^\prime$ into a pair of single generation SM fermions and DM reads,
\begin{align}
& \Gamma_{Z^\prime\to f_i \bar{f_i}}= N_c \frac{g_X^2 M_{Z^\prime}}{24 \pi} \Big[(Q_L^{2}+ Q_R^{2})\Big]\,,
\nonumber\\&
\Gamma_{Z^\prime\to N_1\,N_1}= \frac{g_X^2 M_{Z^\prime}}{24 \pi} \left(1-\frac{4\,M_1^2}{M_{Z^\prime}^2}\right)^\frac{3}{2}\,,
\end{align}
where $N_c=1(3)$ is the color factor for leptons (quarks) and $Q_{L(R)}$ is the B$-$L charge for the left(right handed) SM fermions, and we have taken the limit in which $M_{Z^\prime}$ is much heavier compared to the SM particles. The dominant decay rates of $\phi$ are given by,
\begin{align}
& \Gamma_{\phi\to Z^\prime\,Z^\prime}=\frac{g_X^2}{8 \pi\,r^4}\,\frac{\Mzp^2}{m_\phi}\,\sqrt{1-4\,r^2} \left(12\,r^4-4\,r^2+1\right)\,,
\nonumber\\&
\Gamma_{\phi\to N_i\,N_i}=\frac{g_X^2}{2\pi}\,\left(\frac{M_i}{\Mzp}\right)^2\,m_\phi\,\left(1-\frac{4\,M_i^2}{m_\phi^2}\right)^{3/2}\,,
\end{align}
where $r=M_{Z^\prime}/m_\phi$. For very heavy $Z^\prime$ and DM, the total annihilation cross-section mediated by $Z^\prime$ reads, 
\begin{align}
&\sigma(s)_{\text{SM}\,\text{SM}\to N_1\,N_1}=\frac{170\,g_X^4}{2592\,\pi}\,\frac{s}{\Big[(s-\Mzp^2)^2+\Gamma_{Z^\prime}^2\,\Mzp^2\Big]}
%\nonumber\\&
\left(1-\frac{4\,M_1^2}{s}\right)^{3/2}\,, 
\end{align}
where $\Gamma_{Z^\prime}$ is the total decay width of $Z^\prime$ into all possible final states. 
%%%%%%%%%%%%%%%%%%%%%%%%%%%%%%%%%%%%%
\section{Boltzmann equations for DM analysis}
\label{sec:BEQs}
%%%%%%%%%%%%%%%%%%%%%%%%%%%%%
The coupled BEQs for freeze-in production of the DM, $N_1$, read
\begin{align}
& \frac{dy_\phi}{dz}=-\frac{z}{\mathcal{H}}\,\langle \Gamma_{\phi}\rangle y_{\rm eq}^\phi + \frac{s}{\mathcal{H}}\, \frac{1}{z^2}\,\langle\sigma v\rangle_{\text{SM}\, \text{SM}\to \phi\,\phi}\,\left(y_\text{eq}^\phi\right)^2, \nonumber \\&
\frac{dy_{Z^\prime}}{dz}=- \frac{z}{\mathcal{H}}\langle \Gamma_{Z^\prime}\rangle y_{Z^\prime} +\frac{z}{\mathcal{H}}\langle \Gamma_{\phi \to Z^\prime Z^\prime }\rangle\,\left(y_\text{eq}^\phi-y_{Z^\prime}\right),
\nonumber\\&
\frac{dy_{N_1}}{dz}=\frac{z}{\mathcal{H}}\,\langle \Gamma_{Z^\prime \to N_1N_1}\rangle y_{Z^\prime}+ \frac{z}{\mathcal{H}}\,\langle \Gamma_{\phi\to N_1N_1}\rangle\,y_{\rm eq}^\phi 
+ \frac{s}{\mathcal{H}}\, \frac{1}{z^2}\,\langle\sigma v\rangle_{\text{SM}\, \text{SM}\to N_1 N_1}\,y_\text{eq}^2\,, 
\label{eq:cBEQ}
\end{align}
where $y_i\equiv n_i/s$ is the yield of a certain species $i$, with
\begin{align}
& y_j^\text{eq} = \frac{45}{4\,\pi^4}\,\frac{g_j}{\gss}\,z^2\,K_2[z]\,,    
\end{align}
is the equilibrium yield, with $g_j$ being the degrees of freedom for the corresponding $j$ particle and $z=M_1/T$ is a dimensionless variable. The thermally averaged decay rate $\langle\Gamma_\phi\rangle$ reads,
\begin{align}
& \langle\Gamma_{\phi\to jj}\rangle  = \frac{K_1(z)}{K_2(z)}\times\Gamma_{\phi\to jj}\,,  
\end{align}
with $z=M_1/T$, and $j$ represents the final state particle. Here, $\mathcal{H}=(\pi/3)\,\sqrt{\gs/10}\,\left(T^2/M_P\right)$ is the Hubble parameter for a standard radiation dominated (RD) Universe and $s=\left(2\pi^2/45\right)\,\gss(T)\,T^3$ is the entropy density. The number of relativistic degrees of freedom in the bath corresponding to energy density and entropy density are tracked by $\gs(T)$ and $\gss(T)$, respectively. At temperatures well above the QCD phase transition we have $\gs(T)\simeq\gss(T)\approx 106$.
%%%%%%%%%%%%%%%%%%%%%%%%%%%%%%%%%%%%%%
\section{CP-asymmetry from $N_2$ decay}
\label{sec:CP}
%%%%%%%%%%%%
Since $N_1$ is a long-lived DM, the CP asymmetry generated from $N_2$ decay is given by~\cite{Davidson:2008bu}
\begin{align}\label{eq:cp1}
& \epsilon_{\Delta L} \equiv \frac{\Gamma_{N_2 \to \ell_i\, H } -\Gamma_{N_2 \to \bar\ell_i\, \bar H}}{\Gamma_{N_2 \to \ell_i\, H} + \Gamma_{N_2 \to \bar\ell_i\, \bar H}} \simeq \frac{1}{8\, \pi}\, \frac{1}{(Y_\nu^\dagger\, Y_\nu)_{22}}
\nonumber\\&
\text{Im}\left(Y_\nu^\dagger\, Y_\nu\right)^2_{23}\times \mathcal{F}\left(\frac{M_3^2}{M_2^2}\right)\,,
\end{align}
where
\begin{equation}
   \mathcal{F}(x) \equiv \sqrt{x}\,\left[\frac{1}{1-x}+1-(1+x)\,\log\left(\frac{1+x}{x}\right)\right]\,.
\end{equation}
For $x\gg 1\,,\mathcal{F}\simeq -3/\left(2\,\sqrt{x}\right)$, and Eq.~\eqref{eq:cp1} becomes
\begin{align}
& \epsilon_{\Delta L} \simeq -\frac{3}{16\, \pi}\, \frac{1}{(Y_\nu^\dagger\, Y_\nu)_{22}}\times \left[\text{Im}\left(Y_\nu^\dagger\, Y_\nu\right)^2_{23}\right]\, \frac{M_2}{M_3}\,.
\end{align}
Then 
\begin{equation}
   \epsilon_{\Delta L} \simeq -\frac{3\, \delta_\text{eff}}{16\, \pi}\,\frac{|(Y_\nu)_{23}|^2 \, M_2}{M_3}\,,   
\end{equation}
while the effective CP violating phase is given by
\begin{equation}
   \delta_\text{eff}= \frac{1}{(Y_\nu)_{23}^2}\, \frac{\text{Im}(Y_\nu^\dagger\,Y_\nu)^2_{23}}{(Y_\nu^\dagger\,Y_\nu)_{22}}\,.    
\end{equation}
To connect with the light neutrino mass, we impose the seesaw relation
\begin{equation}
   m_{\nu,3} = \frac{|(Y_\nu)_{23}|^2\, v^2}{M_1}\,,    
\end{equation}
which corresponds to the heaviest left-handed neutrino in the normal hierarchy. This leads to
\begin{equation}
   \epsilon_{\Delta L} \simeq -\frac{3\, \delta_\text{eff}}{16\, \pi}\, \frac{M_2\, m_{\nu,3}}{v^2}\,.
\end{equation}
%%%%%%%%%%%%%%%%%%%%%%%
\section{Boltzmann equations for $N_2$ leptogenesis}
\label{sec:N2lepto}
%%%%%%%%%%%%%%%%%%%%%%%
To track the evolution of RHN number density and the corresponding B$-$L asymmetry with time, we solve the following set of BEQs (along with the first one in Eq.~\eqref{eq:cBEQ}),
\begin{align}
&\frac{dy_{Z^\prime}}{dz}=- \frac{z}{\mathcal{H}}\langle \Gamma_{Z^\prime}\rangle y_{Z^\prime} +\frac{z}{\mathcal{H}}\langle \Gamma_{\phi \to Z^\prime Z^\prime }\rangle\,\left(y_\text{eq}^\phi-y_{Z^\prime}\right)\,, 
\nonumber \\&
\frac{dy_{N_2}}{dz}=\frac{z}{\mathcal{H}}\,\langle\Gamma_{Z^\prime\to N_2\,N_2}\rangle\,y_{Z^\prime}+\frac{z}{\mathcal{H}}\,\langle\Gamma_{\Phi\to N_2\,N_2}\rangle\,y_{\rm eq}^\phi 
%\nonumber\\&
+ \frac{s}{\mathcal{H}} \frac{1}{z^2}\,\langle\sigma v\rangle_{\text{SM}\, \text{SM}\to N_2 N_2}\,\left({y_{N_2}^\text{eq}}^2-y_{N_2}^2\right)
%\nonumber\\&
-\frac{z}{\mathcal{H}}\,\langle\Gamma_{N_2}\rangle\left(y_{N_2}-y_{N_2}^{\rm eq}\right)\,,
\nonumber \\&
\frac{dy_{B-L}}{dz}=\frac{\langle\Gamma_{N_2}\rangle}{\mathcal{H}}z\,\epsilon_{\Delta L}\,\left(y_{N_2}-y_{N_2}^{\rm eq}\right)
-\frac{\langle\Gamma_{N_2}\rangle}{\mathcal{H}}
\frac{z^3}{4}\,\frac{\widetilde m_2}{m_\star}\,K_1(z)\,y_{B-L}\,,
\label{eq:BEQ-N2}
\end{align}
where $z=M_2/T$. Here, $
\widetilde m_2=\left(m_D^\dagger\,m_D\right)_{22}/M_2 \approx m_{\nu,2}^2/M_2$ and and $m_\star\simeq 10^{-3}$ eV is the equilibrium neutrino mass~\cite{Buchmuller:2004nz}. In our scenario, we find, $\widetilde m_2>m_\star$, as a result we are always in the strong washout regime. The thermally averaged $N_2$ decay width is given by
\begin{align}
& \langle \Gamma_{N_2}\rangle=\frac{K_1(M_2/T)}{K_2(M_2/T)}\frac{M_2}{4\pi}\,\frac{m_{\nu,3}\,M_2}{v^2}\,, 
\end{align}
where we have utilized the seesaw relations following normal hierarchy, $m_{\nu,1}\simeq 0$,  $m_{\nu,2} \simeq 0.0086$ eV and $m_{\nu,3}\simeq 0.0506$ eV.
%%%%%%%%%%%%
\section{Renormalization group equations (RGE)}
\label{sec:RGE}
%%%%%%%%%%%%
We extract 1-loop $\beta$-function using {\tt PyR@TE}~\cite{Lyonnet:2013dna,Lyonnet:2016xiz,Sartore:2020gou}. These are given by,
\begin{align}
& \left(4\pi\right)^2\,\beta(g_X)=12\,g_X^3 \nonumber
\\&
\left(4\pi\right)^2\,\beta(Y_\nu)=\frac{3}{2} Y_\nu Y_\nu^{\dagger} Y_\nu+ 3\, \text{Tr}(Y_t^{\dagger} Y_t)\, Y_\nu+
%\\&
\text{Tr}(Y_\nu^{\dagger} Y_\nu)\,Y_\nu- \frac{3}{4} g^2 Y_\nu- 6 g_X^2 Y_\nu
- \frac{9}{4}\,g_2^2\,Y_\nu\, \nonumber 
\\&
\left(4\pi\right)^2\,\beta(\lambda_H)=24\, \lambda_H^{2}
+ \lambda_{\rm mix}^{2}
- 3\, g_1^{2}\,\lambda_H
- 9\, g_2^{2}\,\lambda_H %\nonumber \\
+ \frac{3}{8}\,g_1^{4}
+ \frac{3}{4}\,g_1^{2}\,g_2^{2}
+ \frac{9}{8}\,g_2^{4}+ 12\, \lambda_H\, Y_t^{2}
- 6\, Y_t^{4} \nonumber 
\\&
\left(4\pi\right)^2\,\beta(\lambda_\Phi)=20\,\lambda_H^{2}+ 2\,\lambda_\Phi^{2} \nonumber 
\\&
\left(4\pi\right)^2\,\beta(\lambda_{\rm mix})=12\,\lambda_H\,\lambda_{\rm mix}+8\,\lambda_\Phi\,\lambda_{\rm mix}+4\,\lambda_{\rm mix}^{2}-
%\nonumber\\&
\frac{3}{2}\,g_1^{2}\,\lambda_{\rm mix}-24 g_X^{2} \lambda_{\rm mix}-  \frac{9}{2}\,g_2^{2}\,\lambda_{\rm mix}+ 6\,\lambda_{\rm mix}\,\tr\left(Y_t^{\dagger} Y_t \right)\,, %\nonumber
\end{align}
together with
\begin{align}
& \left(4\pi\right)^2\,\beta(g_1)=\frac{41}{6}\,g_1^3\,, \nonumber 
\\&
\left(4\pi\right)^2\,\beta(g_2)=-\frac{19}{6}\,g_2^3\,,\nonumber
\\&
\left(4\pi\right)^2\,\beta(g_3)=-7\,g_3^3\,, \nonumber
\\&
\left(4\pi\right)^2\,\beta(Y_t)=\frac{3}{2}\,Y_t Y_t^{\dagger}\,Y_t++ 3\,\text{Tr}(Y_t^{\dagger} Y_t)\,Y_t
%\nonumber\\&
-\frac{17}{12}\,g_1^2\,Y_t- \frac{9}{4}\, g_2^2\,Y_t-8\,g_3^2\,Y_t\,.
\end{align}
%%%%%%%%%%%%%%%%%%%%%%%%%%%
\end{widetext}
\vspace{-0.398in}
\bibliographystyle{utphys}
\bibliography{bibliography}

\providecommand{\href}[2]{#2}\begingroup\raggedright\begin{thebibliography}{100}

\bibitem{ParticleDataGroup:2024cfk}
{\bfseries Particle Data Group} Collaboration, S.~Navas {\em et~al.}, ``{Review
  of particle physics},''
  \href{http://dx.doi.org/10.1103/PhysRevD.110.030001}{{\em Phys. Rev. D}
  {\bfseries 110} no.~3, (2024) 030001}.

\bibitem{Planck:2018vyg}
{\bfseries Planck} Collaboration, N.~Aghanim {\em et~al.}, ``{Planck 2018
  results. VI. Cosmological parameters},''
  \href{http://dx.doi.org/10.1051/0004-6361/201833910}{{\em Astron. Astrophys.}
  {\bfseries 641} (2020) A6}, \href{http://arxiv.org/abs/1807.06209}{{\ttfamily
  arXiv:1807.06209 [astro-ph.CO]}}. [Erratum: Astron.Astrophys. 652, C4
  (2021)].

\bibitem{Bertone:2016nfn}
G.~Bertone and D.~Hooper, ``{History of dark matter},''
  \href{http://dx.doi.org/10.1103/RevModPhys.90.045002}{{\em Rev. Mod. Phys.}
  {\bfseries 90} no.~4, (2018) 045002},
  \href{http://arxiv.org/abs/1605.04909}{{\ttfamily arXiv:1605.04909
  [astro-ph.CO]}}.

\bibitem{deSwart:2017heh}
J.~de~Swart, G.~Bertone, and J.~van Dongen, ``{How Dark Matter Came to
  Matter},'' \href{http://dx.doi.org/10.1038/s41550017-0059}{{\em Nature
  Astron.} {\bfseries 1} (2017) 0059},
  \href{http://arxiv.org/abs/1703.00013}{{\ttfamily arXiv:1703.00013
  [astro-ph.CO]}}.

\bibitem{IceCube:2013cdw}
{\bfseries IceCube} Collaboration, M.~G. Aartsen {\em et~al.}, ``{First
  observation of PeV-energy neutrinos with IceCube},''
  \href{http://dx.doi.org/10.1103/PhysRevLett.111.021103}{{\em Phys. Rev.
  Lett.} {\bfseries 111} (2013) 021103},
  \href{http://arxiv.org/abs/1304.5356}{{\ttfamily arXiv:1304.5356
  [astro-ph.HE]}}.

\bibitem{IceCube:2013low}
{\bfseries IceCube} Collaboration, M.~G. Aartsen {\em et~al.}, ``{Evidence for
  High-Energy Extraterrestrial Neutrinos at the IceCube Detector},''
  \href{http://dx.doi.org/10.1126/science.1242856}{{\em Science} {\bfseries
  342} (2013) 1242856}, \href{http://arxiv.org/abs/1311.5238}{{\ttfamily
  arXiv:1311.5238 [astro-ph.HE]}}.

\bibitem{IceCube:2014stg}
{\bfseries IceCube} Collaboration, M.~G. Aartsen {\em et~al.}, ``{Observation
  of High-Energy Astrophysical Neutrinos in Three Years of IceCube Data},''
  \href{http://dx.doi.org/10.1103/PhysRevLett.113.101101}{{\em Phys. Rev.
  Lett.} {\bfseries 113} (2014) 101101},
  \href{http://arxiv.org/abs/1405.5303}{{\ttfamily arXiv:1405.5303
  [astro-ph.HE]}}.

\bibitem{IceCube:2015gsk}
{\bfseries IceCube} Collaboration, M.~G. Aartsen {\em et~al.}, ``{A combined
  maximum-likelihood analysis of the high-energy astrophysical neutrino flux
  measured with IceCube},''
  \href{http://dx.doi.org/10.1088/0004-637X/809/1/98}{{\em Astrophys. J.}
  {\bfseries 809} no.~1, (2015) 98},
  \href{http://arxiv.org/abs/1507.03991}{{\ttfamily arXiv:1507.03991
  [astro-ph.HE]}}.

\bibitem{IceCube:2015qii}
{\bfseries IceCube} Collaboration, M.~G. Aartsen {\em et~al.}, ``{Evidence for
  Astrophysical Muon Neutrinos from the Northern Sky with IceCube},''
  \href{http://dx.doi.org/10.1103/PhysRevLett.115.081102}{{\em Phys. Rev.
  Lett.} {\bfseries 115} no.~8, (2015) 081102},
  \href{http://arxiv.org/abs/1507.04005}{{\ttfamily arXiv:1507.04005
  [astro-ph.HE]}}.

\bibitem{Li:2025tqf}
S.~W. Li, P.~Machado, D.~Naredo-Tuero, and T.~Schwemberger, ``{Clash of the
  Titans: ultra-high energy KM3NeT event versus IceCube data},''
  \href{http://arxiv.org/abs/2502.04508}{{\ttfamily arXiv:2502.04508
  [astro-ph.HE]}}.

\bibitem{Bai:2013nga}
Y.~Bai, R.~Lu, and J.~Salvado, ``{Geometric Compatibility of IceCube TeV-PeV
  Neutrino Excess and its Galactic Dark Matter Origin},''
  \href{http://dx.doi.org/10.1007/JHEP01(2016)161}{{\em JHEP} {\bfseries 01}
  (2016) 161}, \href{http://arxiv.org/abs/1311.5864}{{\ttfamily arXiv:1311.5864
  [hep-ph]}}.

\bibitem{Higaki:2014dwa}
T.~Higaki, R.~Kitano, and R.~Sato, ``{Neutrinoful Universe},''
  \href{http://dx.doi.org/10.1007/JHEP07(2014)044}{{\em JHEP} {\bfseries 07}
  (2014) 044}, \href{http://arxiv.org/abs/1405.0013}{{\ttfamily arXiv:1405.0013
  [hep-ph]}}.

\bibitem{Esmaili:2014rma}
A.~Esmaili, S.~K. Kang, and P.~D. Serpico, ``{IceCube events and decaying dark
  matter: hints and constraints},''
  \href{http://dx.doi.org/10.1088/1475-7516/2014/12/054}{{\em JCAP} {\bfseries
  12} (2014) 054}, \href{http://arxiv.org/abs/1410.5979}{{\ttfamily
  arXiv:1410.5979 [hep-ph]}}.

\bibitem{Murase:2015gea}
K.~Murase, R.~Laha, S.~Ando, and M.~Ahlers, ``{Testing the Dark Matter Scenario
  for PeV Neutrinos Observed in IceCube},''
  \href{http://dx.doi.org/10.1103/PhysRevLett.115.071301}{{\em Phys. Rev.
  Lett.} {\bfseries 115} no.~7, (2015) 071301},
  \href{http://arxiv.org/abs/1503.04663}{{\ttfamily arXiv:1503.04663
  [hep-ph]}}.

\bibitem{Dudas:2018npp}
E.~Dudas, T.~Gherghetta, K.~Kaneta, Y.~Mambrini, and K.~A. Olive, ``{Gravitino
  decay in high scale supersymmetry with R -parity violation},''
  \href{http://dx.doi.org/10.1103/PhysRevD.98.015030}{{\em Phys. Rev. D}
  {\bfseries 98} no.~1, (2018) 015030},
  \href{http://arxiv.org/abs/1805.07342}{{\ttfamily arXiv:1805.07342
  [hep-ph]}}.

\bibitem{Cohen:2016uyg}
T.~Cohen, K.~Murase, N.~L. Rodd, B.~R. Safdi, and Y.~Soreq,
  ``{\ensuremath{\gamma} -ray Constraints on Decaying Dark Matter and
  Implications for IceCube},''
  \href{http://dx.doi.org/10.1103/PhysRevLett.119.021102}{{\em Phys. Rev.
  Lett.} {\bfseries 119} no.~2, (2017) 021102},
  \href{http://arxiv.org/abs/1612.05638}{{\ttfamily arXiv:1612.05638
  [hep-ph]}}.

\bibitem{Rott:2014kfa}
C.~Rott, K.~Kohri, and S.~C. Park, ``{Superheavy dark matter and IceCube
  neutrino signals: Bounds on decaying dark matter},''
  \href{http://dx.doi.org/10.1103/PhysRevD.92.023529}{{\em Phys. Rev. D}
  {\bfseries 92} no.~2, (2015) 023529},
  \href{http://arxiv.org/abs/1408.4575}{{\ttfamily arXiv:1408.4575 [hep-ph]}}.

\bibitem{Arguelles:2022nbl}
C.~A. Arg\"uelles, D.~Delgado, A.~Friedlander, A.~Kheirandish, I.~Safa, A.~C.
  Vincent, and H.~White, ``{Dark Matter decay to neutrinos},''
  \href{http://arxiv.org/abs/2210.01303}{{\ttfamily arXiv:2210.01303
  [hep-ph]}}.

\bibitem{LIGOScientific:2021nrg}
{\bfseries LIGO Scientific, Virgo, KAGRA} Collaboration, R.~Abbott {\em
  et~al.}, ``{Constraints on Cosmic Strings Using Data from the Third Advanced
  LIGO\textendash{}Virgo Observing Run},''
  \href{http://dx.doi.org/10.1103/PhysRevLett.126.241102}{{\em Phys. Rev.
  Lett.} {\bfseries 126} no.~24, (2021) 241102},
  \href{http://arxiv.org/abs/2101.12248}{{\ttfamily arXiv:2101.12248 [gr-qc]}}.

\bibitem{Caldwell:2022qsj}
R.~Caldwell {\em et~al.}, ``{Detection of early-universe gravitational-wave
  signatures and fundamental physics},''
  \href{http://dx.doi.org/10.1007/s10714-022-03027-x}{{\em Gen. Rel. Grav.}
  {\bfseries 54} no.~12, (2022) 156},
  \href{http://arxiv.org/abs/2203.07972}{{\ttfamily arXiv:2203.07972 [gr-qc]}}.

\bibitem{NANOGrav:2023gor}
{\bfseries NANOGrav} Collaboration, G.~Agazie {\em et~al.}, ``{The NANOGrav 15
  yr Data Set: Evidence for a Gravitational-wave Background},''
  \href{http://dx.doi.org/10.3847/2041-8213/acdac6}{{\em Astrophys. J. Lett.}
  {\bfseries 951} no.~1, (2023) L8},
  \href{http://arxiv.org/abs/2306.16213}{{\ttfamily arXiv:2306.16213
  [astro-ph.HE]}}.

\bibitem{NANOGrav:2023hvm}
{\bfseries NANOGrav} Collaboration, A.~Afzal {\em et~al.}, ``{The NANOGrav 15
  yr Data Set: Search for Signals from New Physics},''
  \href{http://dx.doi.org/10.3847/2041-8213/acdc91}{{\em Astrophys. J. Lett.}
  {\bfseries 951} no.~1, (2023) L11},
  \href{http://arxiv.org/abs/2306.16219}{{\ttfamily arXiv:2306.16219
  [astro-ph.HE]}}. [Erratum: Astrophys.J.Lett. 971, L27 (2024), Erratum:
  Astrophys.J. 971, L27 (2024)].

\bibitem{Xu:2023wog}
H.~Xu {\em et~al.}, ``{Searching for the Nano-Hertz Stochastic Gravitational
  Wave Background with the Chinese Pulsar Timing Array Data Release I},''
  \href{http://dx.doi.org/10.1088/1674-4527/acdfa5}{{\em Res. Astron.
  Astrophys.} {\bfseries 23} no.~7, (2023) 075024},
  \href{http://arxiv.org/abs/2306.16216}{{\ttfamily arXiv:2306.16216
  [astro-ph.HE]}}.

\bibitem{EPTA:2023fyk}
{\bfseries EPTA, InPTA:} Collaboration, J.~Antoniadis {\em et~al.}, ``{The
  second data release from the European Pulsar Timing Array - III. Search for
  gravitational wave signals},''
  \href{http://dx.doi.org/10.1051/0004-6361/202346844}{{\em Astron. Astrophys.}
  {\bfseries 678} (2023) A50},
  \href{http://arxiv.org/abs/2306.16214}{{\ttfamily arXiv:2306.16214
  [astro-ph.HE]}}.

\bibitem{LISACosmologyWorkingGroup:2022jok}
{\bfseries LISA Cosmology Working Group} Collaboration, P.~Auclair {\em
  et~al.}, ``{Cosmology with the Laser Interferometer Space Antenna},''
  \href{http://dx.doi.org/10.1007/s41114-023-00045-2}{{\em Living Rev. Rel.}
  {\bfseries 26} no.~1, (2023) 5},
  \href{http://arxiv.org/abs/2204.05434}{{\ttfamily arXiv:2204.05434
  [astro-ph.CO]}}.

\bibitem{Nielsen:1973cs}
H.~B. Nielsen and P.~Olesen, ``{Vortex Line Models for Dual Strings},''
  \href{http://dx.doi.org/10.1016/0550-3213(73)90350-7}{{\em Nucl. Phys. B}
  {\bfseries 61} (1973) 45--61}.

\bibitem{Kibble:1976sj}
T.~W.~B. Kibble, ``{Topology of Cosmic Domains and Strings},''
  \href{http://dx.doi.org/10.1088/0305-4470/9/8/029}{{\em J. Phys. A}
  {\bfseries 9} (1976) 1387--1398}.

\bibitem{Davidson:1978pm}
A.~Davidson, ``{$B−L$ as the fourth color within an $\mathrm{SU}(2)_L \times
  \mathrm{U}(1)_R \times \mathrm{U}(1)$ model},''
  \href{http://dx.doi.org/10.1103/PhysRevD.20.776}{{\em Phys. Rev. D}
  {\bfseries 20} (1979) 776}.

\bibitem{Marshak:1979fm}
R.~E. Marshak and R.~N. Mohapatra, ``{Quark - Lepton Symmetry and B-L as the
  U(1) Generator of the Electroweak Symmetry Group},''
\href{http://dx.doi.org/10.1016/0370-2693(80)90436-0}{{\em Phys. Lett.}
  {\bfseries 91B} (1980) 222--224}.
%%CITATION = PHLTA,91B,222;%%.

\bibitem{Minkowski:1977sc}
P.~Minkowski, ``{$\mu \to e\gamma$ at a Rate of One Out of $10^{9}$ Muon
  Decays?},'' \href{http://dx.doi.org/10.1016/0370-2693(77)90435-X}{{\em Phys.
  Lett. B} {\bfseries 67} (1977) 421--428}.

\bibitem{Yanagida:1979as}
T.~Yanagida, ``{Horizontal gauge symmetry and masses of neutrinos},''
{\em Conf. Proc.} {\bfseries C7902131} (1979) 95--99.
%%CITATION = CONFP,C7902131,95;%%.

\bibitem{Gell-Mann:1979vob}
M.~Gell-Mann, P.~Ramond, and R.~Slansky, ``{Complex Spinors and Unified
  Theories},'' {\em Conf. Proc. C} {\bfseries 790927} (1979) 315--321,
  \href{http://arxiv.org/abs/1306.4669}{{\ttfamily arXiv:1306.4669 [hep-th]}}.

\bibitem{Mohapatra:1979ia}
R.~N. Mohapatra and G.~Senjanovic, ``{Neutrino Mass and Spontaneous Parity
  Nonconservation},'' \href{http://dx.doi.org/10.1103/PhysRevLett.44.912}{{\em
  Phys. Rev. Lett.} {\bfseries 44} (1980) 912}.
[,231(1979)].
%%CITATION = PRLTA,44,912;%%.

\bibitem{Schechter:1980gr}
J.~Schechter and J.~W.~F. Valle, ``{Neutrino Masses in SU(2) x U(1)
  Theories},''
\href{http://dx.doi.org/10.1103/PhysRevD.22.2227}{{\em Phys. Rev.} {\bfseries
  D22} (1980) 2227}.
%%CITATION = PHRVA,D22,2227;%%.

\bibitem{Babul:1987me}
A.~Babul, T.~Piran, and D.~N. Spergel, ``{BOSONIC SUPERCONDUCTING COSMIC
  STRINGS. 1. CLASSICAL FIELD THEORY SOLUTIONS},''
  \href{http://dx.doi.org/10.1016/0370-2693(88)90476-5}{{\em Phys. Lett. B}
  {\bfseries 202} (1988) 307--314}.

\bibitem{Vilenkin:2000jqa}
A.~Vilenkin and E.~P.~S. Shellard, {\em {Cosmic Strings and Other Topological
  Defects}}.
\newblock Cambridge University Press, 7, 2000.

\bibitem{Dror:2019syi}
J.~A. Dror, T.~Hiramatsu, K.~Kohri, H.~Murayama, and G.~White, ``{Testing the
  Seesaw Mechanism and Leptogenesis with Gravitational Waves},''
  \href{http://dx.doi.org/10.1103/PhysRevLett.124.041804}{{\em Phys. Rev.
  Lett.} {\bfseries 124} no.~4, (2020) 041804},
  \href{http://arxiv.org/abs/1908.03227}{{\ttfamily arXiv:1908.03227
  [hep-ph]}}.

\bibitem{Das:2021esm}
A.~Das, P.~S.~B. Dev, Y.~Hosotani, and S.~Mandal, ``{Probing the minimal
  $U(1)_X$ model at future electron-positron colliders via the fermion
  pair-production channel},'' \href{http://arxiv.org/abs/2104.10902}{{\ttfamily
  arXiv:2104.10902 [hep-ph]}}.

\bibitem{Lunardini:2019zob}
C.~Lunardini and Y.~F. Perez-Gonzalez, ``{Dirac and Majorana neutrino
  signatures of primordial black holes},''
  \href{http://dx.doi.org/10.1088/1475-7516/2020/08/014}{{\em JCAP} {\bfseries
  08} (2020) 014}, \href{http://arxiv.org/abs/1910.07864}{{\ttfamily
  arXiv:1910.07864 [hep-ph]}}.

\bibitem{Bernal:2022swt}
N.~Bernal, V.~Mu\~noz Albornoz, S.~Palomares-Ruiz, and P.~Villanueva-Domingo,
  ``{Current and future neutrino limits on the abundance of primordial black
  holes},'' \href{http://dx.doi.org/10.1088/1475-7516/2022/10/068}{{\em JCAP}
  {\bfseries 10} (2022) 068}, \href{http://arxiv.org/abs/2203.14979}{{\ttfamily
  arXiv:2203.14979 [hep-ph]}}.

\bibitem{Wu:2024uxa}
Q.-f. Wu and X.-J. Xu, ``{High-energy and ultra-high-energy neutrinos from
  Primordial Black Holes},''
  \href{http://dx.doi.org/10.1088/1475-7516/2025/02/059}{{\em JCAP} {\bfseries
  02} (2025) 059}, \href{http://arxiv.org/abs/2409.09468}{{\ttfamily
  arXiv:2409.09468 [hep-ph]}}.

\bibitem{Chianese:2024rsn}
M.~Chianese, A.~Boccia, F.~Iocco, G.~Miele, and N.~Saviano, ``{Light burden of
  memory: Constraining primordial black holes with high-energy neutrinos},''
  \href{http://dx.doi.org/10.1103/PhysRevD.111.063036}{{\em Phys. Rev. D}
  {\bfseries 111} no.~6, (2025) 063036},
  \href{http://arxiv.org/abs/2410.07604}{{\ttfamily arXiv:2410.07604
  [astro-ph.HE]}}.

\bibitem{Zantedeschi:2024ram}
M.~Zantedeschi and L.~Visinelli, ``{Ultralight Black Holes as Sources of
  High-Energy Particles},'' \href{http://arxiv.org/abs/2410.07037}{{\ttfamily
  arXiv:2410.07037 [astro-ph.HE]}}.

\bibitem{Baker:2025cff}
M.~J. Baker, J.~Iguaz~Juan, A.~Symons, and A.~Thamm, ``{Explaining the PeV
  Neutrino Fluxes at KM3NeT and IceCube with Quasi-Extremal Primordial Black
  Holes},'' \href{http://arxiv.org/abs/2505.22722}{{\ttfamily arXiv:2505.22722
  [hep-ph]}}.

\bibitem{Fujita:2014hha}
T.~Fujita, K.~Harigaya, and T.~Matsuda, ``{Baryon asymmetry, dark matter, and
  density perturbation from primordial black holes},''
  \href{http://dx.doi.org/10.1103/PhysRevD.89.103501}{{\em Phys. Rev. D}
  {\bfseries 89} (2014) 103501},
  \href{http://arxiv.org/abs/1401.1909}{{\ttfamily arXiv:1401.1909 [hep-ph]}}.

\bibitem{Datta:2020bht}
S.~Datta, A.~Ghosal, and R.~Samanta, ``{Baryogenesis from ultralight primordial
  black holes and strong gravitational waves from cosmic strings},''
  \href{http://dx.doi.org/10.1088/1475-7516/2021/08/021}{{\em JCAP} {\bfseries
  08} (2021) 021}, \href{http://arxiv.org/abs/2012.14981}{{\ttfamily
  arXiv:2012.14981 [hep-ph]}}.

\bibitem{Barman:2021ost}
B.~Barman, D.~Borah, S.~J. Das, and R.~Roshan, ``{Non-thermal origin of
  asymmetric dark matter from inflaton and primordial black holes},''
  \href{http://dx.doi.org/10.1088/1475-7516/2022/03/031}{{\em JCAP} {\bfseries
  03} no.~03, (2022) 031}, \href{http://arxiv.org/abs/2111.08034}{{\ttfamily
  arXiv:2111.08034 [hep-ph]}}.

\bibitem{Bernal:2022pue}
N.~Bernal, C.~S. Fong, Y.~F. Perez-Gonzalez, and J.~Turner, ``{Rescuing
  high-scale leptogenesis using primordial black holes},''
  \href{http://dx.doi.org/10.1103/PhysRevD.106.035019}{{\em Phys. Rev. D}
  {\bfseries 106} no.~3, (2022) 035019},
  \href{http://arxiv.org/abs/2203.08823}{{\ttfamily arXiv:2203.08823
  [hep-ph]}}.

\bibitem{Sawada:1979dis}
O.~Sawada and A.~Sugamoto, eds., {\em {Proceedings: Workshop on the Unified
  Theories and the Baryon Number in the Universe}: {Tsukuba, Japan, February
  13-14, 1979}}.
\newblock Natl.Lab.High Energy Phys., Tsukuba, Japan, 1979.

\bibitem{Mohapatra:1980yp}
R.~N. Mohapatra and G.~Senjanovic, ``{Neutrino Masses and Mixings in Gauge
  Models with Spontaneous Parity Violation},''
  \href{http://dx.doi.org/10.1103/PhysRevD.23.165}{{\em Phys. Rev. D}
  {\bfseries 23} (1981) 165}.

\bibitem{Chianese:2016smc}
M.~Chianese and A.~Merle, ``{A Consistent Theory of Decaying Dark Matter
  Connecting IceCube to the Sesame Street},''
  \href{http://dx.doi.org/10.1088/1475-7516/2017/04/017}{{\em JCAP} {\bfseries
  04} (2017) 017}, \href{http://arxiv.org/abs/1607.05283}{{\ttfamily
  arXiv:1607.05283 [hep-ph]}}.

\bibitem{Hall:2009bx}
L.~J. Hall, K.~Jedamzik, J.~March-Russell, and S.~M. West, ``{Freeze-In
  Production of FIMP Dark Matter},''
  \href{http://dx.doi.org/10.1007/JHEP03(2010)080}{{\em JHEP} {\bfseries 03}
  (2010) 080}, \href{http://arxiv.org/abs/0911.1120}{{\ttfamily arXiv:0911.1120
  [hep-ph]}}.

\bibitem{Bernal:2017kxu}
N.~Bernal, M.~Heikinheimo, T.~Tenkanen, K.~Tuominen, and V.~Vaskonen, ``{The
  Dawn of FIMP Dark Matter: A Review of Models and Constraints},''
  \href{http://dx.doi.org/10.1142/S0217751X1730023X}{{\em Int. J. Mod. Phys. A}
  {\bfseries 32} no.~27, (2017) 1730023},
  \href{http://arxiv.org/abs/1706.07442}{{\ttfamily arXiv:1706.07442
  [hep-ph]}}.

\bibitem{KATRIN:2019yun}
{\bfseries KATRIN} Collaboration, M.~Aker {\em et~al.}, ``{Improved Upper Limit
  on the Neutrino Mass from a Direct Kinematic Method by KATRIN},''
  \href{http://dx.doi.org/10.1103/PhysRevLett.123.221802}{{\em Phys. Rev.
  Lett.} {\bfseries 123} no.~22, (2019) 221802},
  \href{http://arxiv.org/abs/1909.06048}{{\ttfamily arXiv:1909.06048
  [hep-ex]}}.

\bibitem{Dolinski:2019nrj}
M.~J. Dolinski, A.~W.~P. Poon, and W.~Rodejohann, ``{Neutrinoless Double-Beta
  Decay: Status and Prospects},''
  \href{http://dx.doi.org/10.1146/annurev-nucl-101918-023407}{{\em Ann. Rev.
  Nucl. Part. Sci.} {\bfseries 69} (2019) 219--251},
  \href{http://arxiv.org/abs/1902.04097}{{\ttfamily arXiv:1902.04097
  [nucl-ex]}}.

\bibitem{Robens:2015gla}
T.~Robens and T.~Stefaniak, ``{Status of the Higgs Singlet Extension of the
  Standard Model after LHC Run 1},''
  \href{http://dx.doi.org/10.1140/epjc/s10052-015-3323-y}{{\em Eur. Phys. J. C}
  {\bfseries 75} (2015) 104}, \href{http://arxiv.org/abs/1501.02234}{{\ttfamily
  arXiv:1501.02234 [hep-ph]}}.

\bibitem{Chalons:2016jeu}
G.~Chalons, D.~Lopez-Val, T.~Robens, and T.~Stefaniak, ``{The Higgs singlet
  extension at LHC Run 2},'' \href{http://dx.doi.org/10.22323/1.282.1180}{{\em
  PoS} {\bfseries ICHEP2016} (2016) 1180},
  \href{http://arxiv.org/abs/1611.03007}{{\ttfamily arXiv:1611.03007
  [hep-ph]}}.

\bibitem{Das:2022oyx}
A.~Das, S.~Gola, S.~Mandal, and N.~Sinha, ``{Two-component scalar and fermionic
  dark matter candidates in a generic U$(1)_X$ model},''
  \href{http://arxiv.org/abs/2202.01443}{{\ttfamily arXiv:2202.01443
  [hep-ph]}}.

\bibitem{LEPWorkingGroupforHiggsbosonsearches:2003ing}
{\bfseries LEP Working Group for Higgs boson searches, ALEPH, DELPHI, L3, OPAL}
  Collaboration, R.~Barate {\em et~al.}, ``{Search for the standard model Higgs
  boson at LEP},'' \href{http://dx.doi.org/10.1016/S0370-2693(03)00614-2}{{\em
  Phys. Lett. B} {\bfseries 565} (2003) 61--75},
  \href{http://arxiv.org/abs/hep-ex/0306033}{{\ttfamily arXiv:hep-ex/0306033}}.

\bibitem{Wang:2020lkq}
Y.~Wang, M.~Berggren, and J.~List, ``{ILD Benchmark: Search for Extra Scalars
  Produced in Association with a $Z$ boson at $\sqrt{s}=500$ GeV},''
  \href{http://arxiv.org/abs/2005.06265}{{\ttfamily arXiv:2005.06265
  [hep-ex]}}.

\bibitem{CLIC:2018fvx}
{\bfseries CLIC} Collaboration, J.~de~Blas {\em et~al.}, ``{The CLIC Potential
  for New Physics},'' \href{http://dx.doi.org/10.23731/CYRM-2018-003}{{\em CERN
  Yellow Rep. Monogr.} {\bfseries 3} (2018) 1--282},
  \href{http://arxiv.org/abs/1812.02093}{{\ttfamily arXiv:1812.02093
  [hep-ph]}}.

\bibitem{Kaneta:2016vkq}
K.~Kaneta, Z.~Kang, and H.-S. Lee, ``{Right-handed neutrino dark matter under
  the $B − L$ gauge interaction},''
  \href{http://dx.doi.org/10.1007/JHEP02(2017)031}{{\em JHEP} {\bfseries 02}
  (2017) 031}, \href{http://arxiv.org/abs/1606.09317}{{\ttfamily
  arXiv:1606.09317 [hep-ph]}}.

\bibitem{Eijima:2022dec}
S.~Eijima, O.~Seto, and T.~Shimomura, ``{Revisiting sterile neutrino dark
  matter in gauged U(1)B-L model},''
  \href{http://dx.doi.org/10.1103/PhysRevD.106.103513}{{\em Phys. Rev. D}
  {\bfseries 106} no.~10, (2022) 103513},
  \href{http://arxiv.org/abs/2207.01775}{{\ttfamily arXiv:2207.01775
  [hep-ph]}}.

\bibitem{Seto:2024lik}
O.~Seto, T.~Shimomura, and Y.~Uchida, ``{Freeze-in sterile neutrino dark matter
  in feeble gauged $B-L$ model},''
  \href{http://arxiv.org/abs/2404.00654}{{\ttfamily arXiv:2404.00654
  [hep-ph]}}.

\bibitem{ParticleDataGroup:2022pth}
{\bfseries Particle Data Group} Collaboration, R.~L. Workman {\em et~al.},
  ``{Review of Particle Physics},''
  \href{http://dx.doi.org/10.1093/ptep/ptac097}{{\em PTEP} {\bfseries 2022}
  (2022) 083C01}.

\bibitem{Buchmuller:2004nz}
W.~Buchmuller, P.~Di~Bari, and M.~Plumacher, ``{Leptogenesis for
  pedestrians},'' \href{http://dx.doi.org/10.1016/j.aop.2004.02.003}{{\em
  Annals Phys.} {\bfseries 315} (2005) 305--351},
  \href{http://arxiv.org/abs/hep-ph/0401240}{{\ttfamily arXiv:hep-ph/0401240}}.

\bibitem{Kaneta:2019yjn}
K.~Kaneta, Y.~Mambrini, K.~A. Olive, and S.~Verner, ``{Inflation and
  Leptogenesis in High-Scale Supersymmetry},''
  \href{http://dx.doi.org/10.1103/PhysRevD.101.015002}{{\em Phys. Rev. D}
  {\bfseries 101} no.~1, (2020) 015002},
  \href{http://arxiv.org/abs/1911.02463}{{\ttfamily arXiv:1911.02463
  [hep-ph]}}.

\bibitem{Davidson:2002qv}
S.~Davidson and A.~Ibarra, ``{A Lower bound on the right-handed neutrino mass
  from leptogenesis},''
  \href{http://dx.doi.org/10.1016/S0370-2693(02)01735-5}{{\em Phys. Lett. B}
  {\bfseries 535} (2002) 25--32},
  \href{http://arxiv.org/abs/hep-ph/0202239}{{\ttfamily arXiv:hep-ph/0202239}}.

\bibitem{Buttazzo:2013uya}
D.~Buttazzo, G.~Degrassi, P.~P. Giardino, G.~F. Giudice, F.~Sala, A.~Salvio,
  and A.~Strumia, ``Investigating the near-criticality of the higgs boson,''
  \href{http://dx.doi.org/10.1007/JHEP12(2013)089}{{\em JHEP} {\bfseries 12}
  (2013) 089}, \href{http://arxiv.org/abs/1307.3536}{{\ttfamily arXiv:1307.3536
  [hep-ph]}}.

\bibitem{Ringeval:2005kr}
C.~Ringeval, M.~Sakellariadou, and F.~Bouchet, ``{Cosmological evolution of
  cosmic string loops},''
  \href{http://dx.doi.org/10.1088/1475-7516/2007/02/023}{{\em JCAP} {\bfseries
  02} (2007) 023}, \href{http://arxiv.org/abs/astro-ph/0511646}{{\ttfamily
  arXiv:astro-ph/0511646}}.

\bibitem{Blanco-Pillado:2011egf}
J.~J. Blanco-Pillado, K.~D. Olum, and B.~Shlaer, ``{Large parallel cosmic
  string simulations: New results on loop production},''
  \href{http://dx.doi.org/10.1103/PhysRevD.83.083514}{{\em Phys. Rev. D}
  {\bfseries 83} (2011) 083514},
  \href{http://arxiv.org/abs/1101.5173}{{\ttfamily arXiv:1101.5173
  [astro-ph.CO]}}.

\bibitem{Nakayama:2018ptw}
K.~Nakayama and Y.~Tang, ``{Stochastic Gravitational Waves from Particle
  Origin},'' \href{http://dx.doi.org/10.1016/j.physletb.2018.11.023}{{\em Phys.
  Lett. B} {\bfseries 788} (2019) 341--346},
  \href{http://arxiv.org/abs/1810.04975}{{\ttfamily arXiv:1810.04975
  [hep-ph]}}.

\bibitem{Barman:2023ymn}
B.~Barman, N.~Bernal, Y.~Xu, and O.~Zapata, ``{Gravitational wave from graviton
  Bremsstrahlung during reheating},''
  \href{http://dx.doi.org/10.1088/1475-7516/2023/05/019}{{\em JCAP} {\bfseries
  05} (2023) 019}, \href{http://arxiv.org/abs/2301.11345}{{\ttfamily
  arXiv:2301.11345 [hep-ph]}}.

\bibitem{Barman:2023rpg}
B.~Barman, N.~Bernal, Y.~Xu, and O.~Zapata, ``{Bremsstrahlung-induced
  gravitational waves in monomial potentials during reheating},''
  \href{http://dx.doi.org/10.1103/PhysRevD.108.083524}{{\em Phys. Rev. D}
  {\bfseries 108} no.~8, (2023) 083524},
  \href{http://arxiv.org/abs/2305.16388}{{\ttfamily arXiv:2305.16388
  [hep-ph]}}.

\bibitem{Choi:2024acs}
K.-Y. Choi, E.~Lkhagvadorj, and S.~Mahapatra, ``{Gravitational wave sourced by
  decay of massive particle from primordial black hole evaporation},''
  \href{http://dx.doi.org/10.1088/1475-7516/2024/07/064}{{\em JCAP} {\bfseries
  07} (2024) 064}, \href{http://arxiv.org/abs/2403.15269}{{\ttfamily
  arXiv:2403.15269 [hep-ph]}}.

\bibitem{Ghoshal:2022kqp}
A.~Ghoshal, R.~Samanta, and G.~White, ``{Bremsstrahlung high-frequency
  gravitational wave signatures of high-scale nonthermal leptogenesis},''
  \href{http://dx.doi.org/10.1103/PhysRevD.108.035019}{{\em Phys. Rev. D}
  {\bfseries 108} no.~3, (2023) 035019},
  \href{http://arxiv.org/abs/2211.10433}{{\ttfamily arXiv:2211.10433
  [hep-ph]}}.

\bibitem{Datta:2024tne}
A.~Datta and A.~Sil, ``{Probing Leptogenesis through Gravitational Waves},''
  \href{http://arxiv.org/abs/2410.01900}{{\ttfamily arXiv:2410.01900
  [hep-ph]}}.

\bibitem{Cata:2016epa}
O.~Cat\`a, A.~Ibarra, and S.~Ingenh\"utt, ``{Dark matter decay through gravity
  portals},'' \href{http://dx.doi.org/10.1103/PhysRevD.95.035011}{{\em Phys.
  Rev. D} {\bfseries 95} no.~3, (2017) 035011},
  \href{http://arxiv.org/abs/1611.00725}{{\ttfamily arXiv:1611.00725
  [hep-ph]}}.

\bibitem{Cata:2017jar}
O.~Cat\`a, A.~Ibarra, and S.~Ingenh\"utt, ``{Sharp spectral features from light
  dark matter decay via gravity portals},''
  \href{http://dx.doi.org/10.1088/1475-7516/2017/11/044}{{\em JCAP} {\bfseries
  11} (2017) 044}, \href{http://arxiv.org/abs/1707.08480}{{\ttfamily
  arXiv:1707.08480 [hep-ph]}}.

\bibitem{Vilenkin:1981bx}
A.~Vilenkin, ``{Gravitational radiation from cosmic strings},''
  \href{http://dx.doi.org/10.1016/0370-2693(81)91144-8}{{\em Phys. Lett. B}
  {\bfseries 107} (1981) 47--50}.

\bibitem{Vachaspati:1984gt}
T.~Vachaspati and A.~Vilenkin, ``{Gravitational Radiation from Cosmic
  Strings},'' \href{http://dx.doi.org/10.1103/PhysRevD.31.3052}{{\em Phys. Rev.
  D} {\bfseries 31} (1985) 3052}.

\bibitem{Blanco-Pillado:2013qja}
J.~J. Blanco-Pillado, K.~D. Olum, and B.~Shlaer, ``{The number of cosmic string
  loops},'' \href{http://dx.doi.org/10.1103/PhysRevD.89.023512}{{\em Phys. Rev.
  D} {\bfseries 89} no.~2, (2014) 023512},
  \href{http://arxiv.org/abs/1309.6637}{{\ttfamily arXiv:1309.6637
  [astro-ph.CO]}}.

\bibitem{Blanco-Pillado:2017oxo}
J.~J. Blanco-Pillado and K.~D. Olum, ``{Stochastic gravitational wave
  background from smoothed cosmic string loops},''
  \href{http://dx.doi.org/10.1103/PhysRevD.96.104046}{{\em Phys. Rev. D}
  {\bfseries 96} no.~10, (2017) 104046},
  \href{http://arxiv.org/abs/1709.02693}{{\ttfamily arXiv:1709.02693
  [astro-ph.CO]}}.

\bibitem{Damour:2001bk}
T.~Damour and A.~Vilenkin, ``{Gravitational wave bursts from cusps and kinks on
  cosmic strings},'' \href{http://dx.doi.org/10.1103/PhysRevD.64.064008}{{\em
  Phys. Rev. D} {\bfseries 64} (2001) 064008},
  \href{http://arxiv.org/abs/gr-qc/0104026}{{\ttfamily arXiv:gr-qc/0104026}}.

\bibitem{Gouttenoire:2019kij}
Y.~Gouttenoire, G.~Servant, and P.~Simakachorn, ``{Beyond the Standard Models
  with Cosmic Strings},''
  \href{http://dx.doi.org/10.1088/1475-7516/2020/07/032}{{\em JCAP} {\bfseries
  07} (2020) 032}, \href{http://arxiv.org/abs/1912.02569}{{\ttfamily
  arXiv:1912.02569 [hep-ph]}}.

\bibitem{Martins:1996jp}
C.~J. A.~P. Martins and E.~P.~S. Shellard, ``{Quantitative string evolution},''
  \href{http://dx.doi.org/10.1103/PhysRevD.54.2535}{{\em Phys. Rev. D}
  {\bfseries 54} (1996) 2535--2556},
  \href{http://arxiv.org/abs/hep-ph/9602271}{{\ttfamily arXiv:hep-ph/9602271}}.

\bibitem{Martins:2000cs}
C.~J. A.~P. Martins and E.~P.~S. Shellard, ``{Extending the velocity dependent
  one scale string evolution model},''
  \href{http://dx.doi.org/10.1103/PhysRevD.65.043514}{{\em Phys. Rev. D}
  {\bfseries 65} (2002) 043514},
  \href{http://arxiv.org/abs/hep-ph/0003298}{{\ttfamily arXiv:hep-ph/0003298}}.

\bibitem{Auclair:2019wcv}
P.~Auclair {\em et~al.}, ``{Probing the gravitational wave background from
  cosmic strings with LISA},''
  \href{http://dx.doi.org/10.1088/1475-7516/2020/04/034}{{\em JCAP} {\bfseries
  04} (2020) 034}, \href{http://arxiv.org/abs/1909.00819}{{\ttfamily
  arXiv:1909.00819 [astro-ph.CO]}}.

\bibitem{Vilenkin:1991zk}
A.~Vilenkin, ``{Cosmic string dynamics with friction},''
  \href{http://dx.doi.org/10.1103/PhysRevD.43.1060}{{\em Phys. Rev. D}
  {\bfseries 43} (1991) 1060--1062}.

\bibitem{Charnock:2016nzm}
T.~Charnock, A.~Avgoustidis, E.~J. Copeland, and A.~Moss, ``{CMB constraints on
  cosmic strings and superstrings},''
  \href{http://dx.doi.org/10.1103/PhysRevD.93.123503}{{\em Phys. Rev. D}
  {\bfseries 93} no.~12, (2016) 123503},
  \href{http://arxiv.org/abs/1603.01275}{{\ttfamily arXiv:1603.01275
  [astro-ph.CO]}}.

\bibitem{Kume:2024adn}
J.~Kume and M.~Hindmarsh, ``{Revised bounds on local cosmic strings from
  NANOGrav observations},''
  \href{http://dx.doi.org/10.1088/1475-7516/2024/12/001}{{\em JCAP} {\bfseries
  12} (2024) 001}, \href{http://arxiv.org/abs/2404.02705}{{\ttfamily
  arXiv:2404.02705 [astro-ph.CO]}}.

\bibitem{Lizarraga:2016onn}
J.~Lizarraga, J.~Urrestilla, D.~Daverio, M.~Hindmarsh, and M.~Kunz, ``{New CMB
  constraints for Abelian Higgs cosmic strings},''
  \href{http://dx.doi.org/10.1088/1475-7516/2016/10/042}{{\em JCAP} {\bfseries
  10} (2016) 042}, \href{http://arxiv.org/abs/1609.03386}{{\ttfamily
  arXiv:1609.03386 [astro-ph.CO]}}.

\bibitem{Crowder:2005nr}
J.~Crowder and N.~J. Cornish, ``{Beyond LISA: Exploring future gravitational
  wave missions},'' \href{http://dx.doi.org/10.1103/PhysRevD.72.083005}{{\em
  Phys. Rev. D} {\bfseries 72} (2005) 083005},
  \href{http://arxiv.org/abs/gr-qc/0506015}{{\ttfamily arXiv:gr-qc/0506015}}.

\bibitem{Corbin:2005ny}
V.~Corbin and N.~J. Cornish, ``{Detecting the cosmic gravitational wave
  background with the big bang observer},''
  \href{http://dx.doi.org/10.1088/0264-9381/23/7/014}{{\em Class. Quant. Grav.}
  {\bfseries 23} (2006) 2435--2446},
  \href{http://arxiv.org/abs/gr-qc/0512039}{{\ttfamily arXiv:gr-qc/0512039}}.

\bibitem{Seto:2001qf}
N.~Seto, S.~Kawamura, and T.~Nakamura, ``{Possibility of direct measurement of
  the acceleration of the universe using 0.1-Hz band laser interferometer
  gravitational wave antenna in space},''
  \href{http://dx.doi.org/10.1103/PhysRevLett.87.221103}{{\em Phys. Rev. Lett.}
  {\bfseries 87} (2001) 221103},
  \href{http://arxiv.org/abs/astro-ph/0108011}{{\ttfamily
  arXiv:astro-ph/0108011}}.

\bibitem{Kudoh:2005as}
H.~Kudoh, A.~Taruya, T.~Hiramatsu, and Y.~Himemoto, ``{Detecting a
  gravitational-wave background with next-generation space interferometers},''
  \href{http://dx.doi.org/10.1103/PhysRevD.73.064006}{{\em Phys. Rev. D}
  {\bfseries 73} (2006) 064006},
  \href{http://arxiv.org/abs/gr-qc/0511145}{{\ttfamily arXiv:gr-qc/0511145}}.

\bibitem{LISA:2017pwj}
{\bfseries LISA} Collaboration, P.~Amaro-Seoane {\em et~al.}, ``{Laser
  Interferometer Space Antenna},''
  \href{http://arxiv.org/abs/1702.00786}{{\ttfamily arXiv:1702.00786
  [astro-ph.IM]}}.

\bibitem{Reitze:2019iox}
D.~Reitze {\em et~al.}, ``{Cosmic Explorer: The U.S. Contribution to
  Gravitational-Wave Astronomy beyond LIGO},'' {\em Bull. Am. Astron. Soc.}
  {\bfseries 51} no.~7, (2019) 035,
  \href{http://arxiv.org/abs/1907.04833}{{\ttfamily arXiv:1907.04833
  [astro-ph.IM]}}.

\bibitem{Hild:2010id}
S.~Hild {\em et~al.}, ``{Sensitivity Studies for Third-Generation Gravitational
  Wave Observatories},''
  \href{http://dx.doi.org/10.1088/0264-9381/28/9/094013}{{\em Class. Quant.
  Grav.} {\bfseries 28} (2011) 094013},
  \href{http://arxiv.org/abs/1012.0908}{{\ttfamily arXiv:1012.0908 [gr-qc]}}.

\bibitem{Punturo:2010zz}
M.~Punturo {\em et~al.}, ``{The Einstein Telescope: A third-generation
  gravitational wave observatory},''
  \href{http://dx.doi.org/10.1088/0264-9381/27/19/194002}{{\em Class. Quant.
  Grav.} {\bfseries 27} (2010) 194002}.

\bibitem{Sathyaprakash:2012jk}
B.~Sathyaprakash {\em et~al.}, ``{Scientific Objectives of Einstein
  Telescope},'' \href{http://dx.doi.org/10.1088/0264-9381/29/12/124013}{{\em
  Class. Quant. Grav.} {\bfseries 29} (2012) 124013},
  \href{http://arxiv.org/abs/1206.0331}{{\ttfamily arXiv:1206.0331 [gr-qc]}}.
  [Erratum: Class.Quant.Grav. 30, 079501 (2013)].

\bibitem{Maggiore:2019uih}
M.~Maggiore {\em et~al.}, ``{Science Case for the Einstein Telescope},''
  \href{http://dx.doi.org/10.1088/1475-7516/2020/03/050}{{\em JCAP} {\bfseries
  03} (2020) 050}, \href{http://arxiv.org/abs/1912.02622}{{\ttfamily
  arXiv:1912.02622 [astro-ph.CO]}}.

\bibitem{Davidson:2008bu}
S.~Davidson, E.~Nardi, and Y.~Nir, ``{Leptogenesis},''
  \href{http://dx.doi.org/10.1016/j.physrep.2008.06.002}{{\em Phys. Rept.}
  {\bfseries 466} (2008) 105--177},
  \href{http://arxiv.org/abs/0802.2962}{{\ttfamily arXiv:0802.2962 [hep-ph]}}.

\bibitem{Lyonnet:2013dna}
F.~Lyonnet, I.~Schienbein, F.~Staub, and A.~Wingerter, ``{PyR@TE:
  Renormalization Group Equations for General Gauge Theories},''
  \href{http://dx.doi.org/10.1016/j.cpc.2013.12.002}{{\em Comput. Phys.
  Commun.} {\bfseries 185} (2014) 1130--1152},
  \href{http://arxiv.org/abs/1309.7030}{{\ttfamily arXiv:1309.7030 [hep-ph]}}.

\bibitem{Lyonnet:2016xiz}
F.~Lyonnet and I.~Schienbein, ``{PyR@TE 2: A Python tool for computing RGEs at
  two-loop},'' \href{http://dx.doi.org/10.1016/j.cpc.2016.12.003}{{\em Comput.
  Phys. Commun.} {\bfseries 213} (2017) 181--196},
  \href{http://arxiv.org/abs/1608.07274}{{\ttfamily arXiv:1608.07274
  [hep-ph]}}.

\bibitem{Sartore:2020gou}
L.~Sartore and I.~Schienbein, ``{PyR@TE 3},''
  \href{http://dx.doi.org/10.1016/j.cpc.2020.107819}{{\em Comput. Phys.
  Commun.} {\bfseries 261} (2021) 107819},
  \href{http://arxiv.org/abs/2007.12700}{{\ttfamily arXiv:2007.12700
  [hep-ph]}}.

\end{thebibliography}\endgroup
%%%%%%%%%%%%%%%%%%%%%%%%%%%
%%%%%%%%%%%%%%%%%%%%%%%%%%%
\end{document}